\documentclass[a4paper,12pt]{article}
\usepackage{jheppub,esint,psfrag}
\usepackage[utf8]{inputenc}

\usepackage{graphicx}
\usepackage{caption}
\usepackage{subcaption}
\usepackage{placeins}
\usepackage{amsmath}
\usepackage{amsfonts}
\usepackage{mathtools}
\usepackage{tikz}

\usepackage{epsfig,amssymb,amsmath,psfrag,rotate,color,wasysym,xcolor}
\usepackage[bbgreekl]{mathbbol}

\allowdisplaybreaks 
\newcommand{\insertfig}[2]{\includegraphics[width=#1cm]{#2}}

\DeclareSymbolFontAlphabet{\mathbbm}{bbold}
\DeclareSymbolFontAlphabet{\mathbb}{AMSb}%

\def\XXint#1#2#3{{\setbox0=\hbox{$#1{#2#3}{\int}$ }
\vcenter{\hbox{$#2#3$ }}\kern-.6\wd0}}

\def \be  {\begin{equation}}
\def \ee  {\end{equation}}
\def \ba  {\begin{eqnarray}}
\def \ea  {\end{eqnarray}}
\def \baa {\begin{eqnarray*}}
\def \eaa {\end{eqnarray*}}
\newcommand{\ep}{\varepsilon}
\def \lab #1 {\label{#1}}

\newcommand\re[1]{(\ref{#1})}
\def\d{\hbox{{d}\kern-.20em\hbox{l}}}

\def \matrix #1 {\left(\begin{array}{cc} #1 \end{array}\right)}

\def \e  {\mathop{\rm e}\nolimits}

\newcommand{\bit}[1]{\mbox{\boldmath$#1$}}
\def\1{\hbox{{1}\kern-.25em\hbox{l}}}

\newcommand{\ft}[2]{{\textstyle\frac{#1}{#2}}}

\makeatletter

\input pdf-trans
\newbox\qbox
\def\usecolor#1{\csname\string\color@#1\endcsname\space}
\newcommand\bordercolor[1]{\colsplit{1}{#1}}
\newcommand\fillcolor[1]{\colsplit{0}{#1}}
\newcommand\outline[1]{\leavevmode%
  \def\maltext{#1}%
  \setbox\qbox=\hbox{\maltext}%
  \boxgs{Q q 2 Tr \thickness\space w \fillcol\space \bordercol\space}{}%
  \copy\qbox%
}
\makeatother
\newcommand\colsplit[2]{\colorlet{tmpcolor}{#2}\edef\tmp{\usecolor{tmpcolor}}%
  \def\tmpB{}\expandafter\colsplithelp\tmp\relax%
  \ifnum0=#1\relax\edef\fillcol{\tmpB}\else\edef\bordercol{\tmpC}\fi}
\def\colsplithelp#1#2 #3\relax{%
  \edef\tmpB{\tmpB#1#2 }%
  \ifnum `#1>`9\relax\def\tmpC{#3}\else\colsplithelp#3\relax\fi
}
\bordercolor{black}
\fillcolor{white}
\def\thickness{.3}

\def\1{\mathbbm{1}}

\title{Tropical regions of near mass-shell pentabox}

\author[a]{A.V.~Belitsky,}
\author[b]{V.A. Smirnov}
\affiliation[a]{Department of Physics, Arizona State University, Tempe, AZ 85287-1504, USA}  
\affiliation[b]{Skobeltsyn Institute of Nuclear Physics, Moscow State University, 119992 Moscow, Russia\\
Moscow Center for Fundamental and Applied Mathematics, 119992 Moscow, Russia}
    
\abstract
{Coulomb branch amplitudes of maximally supersymmetric Yang-Mills theory display infrared properties different
from their conformal counterparts. While the four-leg amplitude is known to very high perturbative orders, amplitudes
of higher multiplicity fall into an uncharted territory starting already from two loops. The reason for this is that they are 
not easily amenable to traditional techniques like canonical differential equations due to the uncontrolled swelling of 
solutions to integration-by-parts identities. In this paper, we break the barrier for the five-leg amplitude using a technique 
based on the analysis of Newton polytopes corresponding to Feynman/Schwinger integrands and their tropical geometry. 
Specifically, we analytically evaluate the near mass-shell limit of the off-shell pentabox in terms of Goncharov polylogarithms.}

\begin{document}

\maketitle
\flushbottom
\setcounter{footnote} 0

\section{Introduction}

\begin{figure}[t]
\begin{center}
\mbox{
\begin{picture}(0,115)(230,0)
\put(0,0){\insertfig{16}{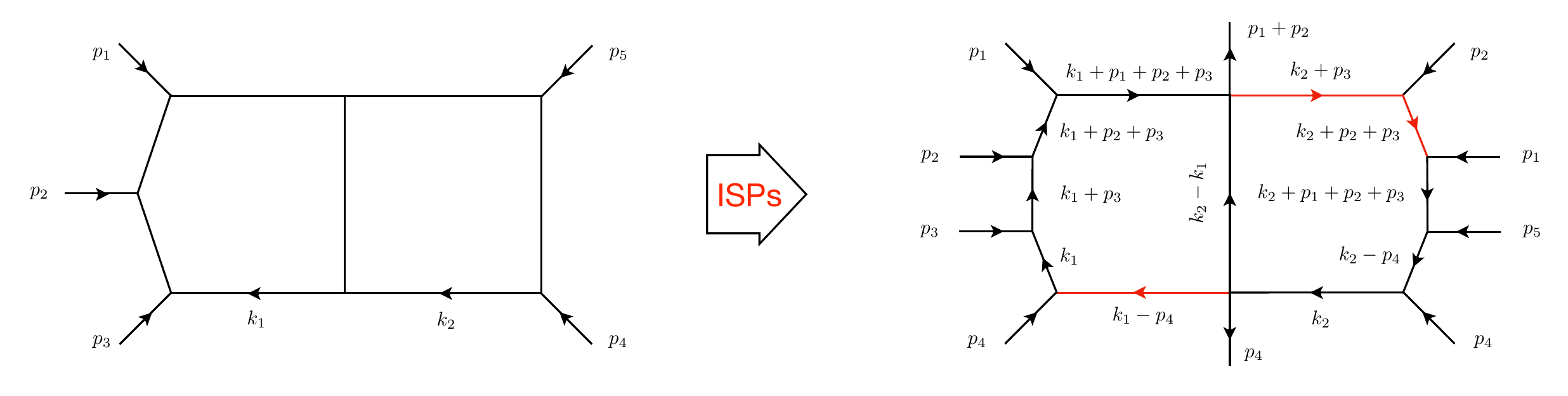}}
\end{picture}
}
\end{center}
\caption{\label{PentaBoxFig} The pentabox (left panel) belongs to a family of Feynman integrals with irreducible scalar products (ISPs)
shown in the right panel.}
\end{figure}

The Coulomb branch \cite{Selivanov:1999ie,Alday:2009zm,Boels:2010mj,Craig:2011ws} of maximally supersymmetric 
Yang-Mills theory, aka $\mathcal{N} = 4$ sYM, is a laboratory to study off-shell scattering of massless particles in a 
controlled gauge-invariant environment. Recent years have taught us 
\cite{Caron-Huot:2021usw,Bork:2022vat,Belitsky:2022itf,Belitsky:2023ssv,Belitsky:2024agy,Belitsky:2024dcf,Belitsky:2025bez}
that the near mass-shell behavior of infrared-sensitive observables in this theory deviates from its conformal point, i.e., the origin 
of the moduli space. Already, leading infrared logarithms of particles' virtualities, which replace poles in the parameter of dimensional 
regularization, are accompanied by different functions of the Yang-Mills coupling constant, or the 't Hooft coupling of the planar limit.
They are given by the cusp \cite{Polyakov:1980ca,Korchemsky:1987wg} and octagon 
\cite{Coronado:2018cxj,Belitsky:2019fan,Belitsky:2020qzm} anomalous dimensions, respectively. Finite remainders differ as well
\cite{Belitsky:2024agy,Belitsky:2024dcf}. In both cases, not only the infrared logarithms exponentiate order-by-order in the coupling, 
but also these finite pieces. Amplitudes on the conformal branch of the theory display a compelling iterative structure 
\cite{Anastasiou:2003kj,Bern:2005iz}, which predicts their all-order form up to additive contributions depending on conformal cross-ratios 
built from particles' momenta. Is this property preserved on the Coulomb branch? Beyond four external legs 
\cite{Caron-Huot:2021usw,He:2025vqt}, this question cannot even be addressed since only one family of Feynman integrals is known 
at two-loop order in the planar limit, i.e., the double box \cite{He:2022ctv,Belitsky:2023gba}. The latter is an indispensable ingredient in 
the basis of the integral decomposition of multileg amplitudes; however, other, more complicated families are currently unknown. For the 
five off-shell external legs \cite{Bork:2022vat,Belitsky:2024rwv}, the pentabox, shown in Fig.\ \ref{PentaBoxFig} (left panel), is needed. 
This is the subject of this work.

Calculating the off-shell pentabox is not an easy feat. The jump in complexity from the double box graph is staggering. While the latter
depends on three Mandelstam invariants and external off-shellness\footnote{We choose it to be the same for all legs.}, the former is a 
function of five Mandelstam-like variables and the off-shellness. We will dub the off-shellness as the `small' invariant, while all the rest as 
`large', since we are interested in the infrared behavior when the former goes to zero. The double box was solved with the conventional 
application of the method of differential equations \cite{Kotikov:1990kg,Gehrmann:1999as,Henn:2013pwa}, and required sensible 
computational resources and time consumption for the successful completion of its analysis \cite{He:2022ctv,Belitsky:2023gba}. In 
particular, solutions to integration-by-parts identities, which are at the heart of the formalism, could be reached in a matter of days in the fully 
analytical approach, or mere hours with modular arithmetic and subsequent rational reconstruction. The five-leg case did not yield to this 
technique. Neither the fully analytical nor the modular techniques for the required integration-by-parts reductions could be handled on a 
supercomputer with 2T of memory\footnote{2T is a user limit on an undedicated cluster.} \cite{Belitsky:2024jhe}.

Were the emerging Feynman integrals finite, the problem could be, at least hypothetically, approached by relying on iterative integration 
techniques of corresponding Feynman/Schwinger parametric integrals. Of course, these have limitations on their own. In general circumstances, 
if parametric integrands are not linearly reducible, the functional space would have to be enlarged from Goncharov polylogarithms (GPLs) to 
elliptical (or higher-order curve) functions and studied on a case-by-case basis. However, if we know a priori that a given Feynman integral can 
be expressed solely in terms of GPLs, how do we go about ubiquitous divergences of loop-momentum integrals?

The predicament spelled out in the last sentence can be resolved as follows. The off-shellness serves as a regulator of infrared divergences. 
In the near mass-shell limit, the latter manifest themselves as its logarithms of decreasing order, starting with the power twice the perturbative 
order. We anticipate that the functional space of their coefficients, and a finite remainder, is spanned by GPLs, which can be tackled with 
computer codes like {\tt HyperInt} \cite{Panzer:2014gra}. So first, we have to recast these coefficients in terms of some (reduced) Feynman 
integrals. This is accomplished relying on the Method of Regions (MofR) \cite{Beneke:1997zp}. However, since the latter `splits up' the 
loop-momentum space into components possessing different scaling with respect to the `small' off-shellness parameter, it inevitably introduces an 
intermediate regularization parameter to make individual momentum components well defined. In a wide range of applications, dimensional 
regularization in $D = 4 - 2 \ep$ suffices for this purpose\footnote{Sometimes, when it is not enough, an additional, analytic regulator is 
employed.}. At the end, when one splices various contributions together, the dependence on $\ep$ vanishes.

MofR provides the sought-after reduced Feynman integrals independent of the `small' off-shell\-ness scale. However, these now depend on
$\ep$ in addition to the `large' external momentum invariants. They develop poles in $\ep$ and obstruct the immediate use of, say, 
{\tt HyperInt}. There are, of course, several well-known methods for resolving the pole structure of dimensionally regularized Feynman 
integrals, such as the Mellin-Barnes contour-integral representation \cite{Smirnov:1999gc,Tausk:1999vh,Czakon:2005rk,Smirnov:2009up}, 
various sector decompositions \cite{Binoth:2000ps,Bogner:2007cr,Kaneko:2009qx,Smirnov:2021rhf}, and the Nilsson-Passare analytic 
continuation \cite{Nilsson,Berkesch}. Unfortunately, the Mellin-Barnes technique is helpless for analytical calculations in problems with many 
Lorentz invariants. The other two are more promising. 

Let us briefly recall them by starting with sector decompositions. According to this procedure, a given Feynman/Schwinger parametric integral 
is broken into the so-called sectors, where for each of them, a new set of integration variables is introduced in such a manner that their integrands 
are represented as analytic functions accompanied by products of sector variables raised to powers linearly dependent on the regularization 
parameter. The immediate consequence of this parametrization is that the analysis of poles in the regularization parameter reduces to the power 
counting of sector variables. Sector decompositions were used in the past to prove basic results on analytic \cite{Speer:1968qxh,Speer:1971fub} 
and dimensional \cite{Breitenlohner:1975hg,Breitenlohner:1976te} renormalization. From the point of view of the original Feynman integrals, they 
were endowed with a clear graph-theoretical meaning by associating sectors with one-particle-irreducible subgraphs of a given graph. A few 
decades down the road, recursive approaches to sector decompositions were devised in Refs.\ \cite{Binoth:2000ps,Bogner:2007cr,Smirnov:2021rhf} 
and successfully applied to the numerical evaluation of Feynman integrals. A modern variant of nonrecursive sector decompositions based on the 
geometry of polytopes was suggested in Ref.~\cite{Kaneko:2009qx}. However, repeating the fate of the Melin-Barnes technique, the sector 
decompositions are inappropriate for analytical evaluations because individual sector contributions become analytically more involved than the 
initial integral they start with. For example, in the case of the strategy of Ref.~\cite{Kaneko:2009qx}, the sector decomposition breaks the linear 
reducibility of parametric integrands by virtue of triangulation of the integration domain and accompanying changes of variables.

Finally, the Nilsson-Passare analytic continuation allows one to extract the poles in the parameter of dimensional regularization by enlarging the
region of analyticity for the corresponding parametric integrands. In retrospect, it echoes the procedure due to 't Hooft and Veltman for analytic 
continuation of dimensionally-regulated Feynman integrals directly in the momentum space \cite{tHooft:1972tcz}. Both of these are based on  
integration by parts (IBP). However, the Nilsson-Passare process possesses two flaws. First, upon the extraction of poles, it does not reduce the 
dimensionality of the accompanying remaining integral. Second, it proliferates the number of integrals after each IBP step. So it quickly becomes
impractical due to the very fast growth of the number of finite integrals at the end of the procedure to be fed, for instance, into {\tt HyperInt}. 

In this work, we will employ and refine a novel strategy recently advocated in Ref.\ \cite{Salvatori:2024nva}. This will allow us to analytically
evaluate contributions from all the regions appearing within MofR. Let us emphasize that this modified strategy can, in principle, be applied 
to any parametric integral, not only to `reduced' region integrals presently considered.

Our subsequent presentation is organized as follows. First, we present the pentabox integrals with ISPs that are the main objects of our 
consideration. In Sect.\ \ref{MofRsection}, we apply MofR in the limit of the off-shellness tending to zero and provide a set of reduced 
regularized Feynman integrals that need to be evaluated. In Sects.\ \ref{TropicsSection} and \ref{LocFiniteSection}, we offer a synopsis of 
the method of Ref.\ \cite{Salvatori:2024nva}. Since the technique is quite new and was previously practically applied in realistic circumstances
only once, i.e., to the single-scale Sudakov form factor in Ref.\ \cite{Belitsky:2025bez}, for the reader's convenience, we will provide there 
an extensive discussion of a sample multiscale integral with various options to tackle it. After that, in Sect.\ \ref{NonsimplicialFanSection}, we 
turn to other regions, continuing with a more involved example that possesses new features compared to the previous one, and which are 
worth emphasizing. In Sect.\ \ref{LinearRedSection}, we address the question of breaking of linear reducibility of integrands and how to tackle 
it. We summarize our results in Sect.\ \ref{ResultsSection}, and finally provide a conclusion. An appendix contains the explicit list of regions' 
integrands obtained within MofR.

\section{Tripentagon family}

The tripentagon family of Feynman integrals is shown in the right panel of Fig.\ \ref{PentaBoxFig}. Anticipating the necessity of dimensional
regularization for MofR, as was alluded to in the introduction, we define them as
\begin{align}
\label{PentaboxFI}
G_{a_1 \dots a_{11}} \equiv \left(\mu^{2}  \e^{\gamma_{\rm E}} \right)^{2 \ep} 
\int \frac{d^D k_1}{i \pi^{D/2}} \frac{d^D k_2}{i \pi^{D/2}} \prod_{j = 1}^{11} D_j^{- a_j}
\, , 
\end{align}
with eight internal propagators ($a_{1, \dots, 8} \geq 0$) and three ISPs ($a_{9,10,11} \leq 0$)
\begin{align}
\label{Propragators} 
&
D_1 = - k_1^2 
\, , \quad
&&
D_2 = - (k_1 + p_3)^2 
\, , \quad
&&
D_3 = - (k_1 + p_2 + p_3)^2
\, , \nonumber\\
&
D_4 = - (k_1 + p_1 + p_2 + p_3)^2
\, , \quad
&&
D_5 = - (k_2 - k_1)^2
\, , \quad
&&
D_6 = - k_2^2
\, , \nonumber\\
&
D_7 = - (k_2 - p_4)^2
\, , \quad
&&
D_8 = -(k_2 + p_1 + p_2 + p_3)^2
\, , \quad
&&
&&
\\
&
D_9 = - (k_1 - p_4)^2
\, , \quad
&&
D_{10} = - (k_2 + p_3)^2
\, , \quad
&&
D_{11} = - (k_2 + p_2 + p_3)^2
\, . \nonumber
\end{align}
These integrals are highly nontrivial functions of the five `large' Mandelstam-like variables $s_{jj+1}$ and the `small' off-shellness $m$
\begin{align}
s_{jj+1} = - (p_j + p_{j+1})^2
\, , \qquad 
m^2 = - p_j^2
\, .
\end{align}
The latter is chosen to be the same for all legs. It is tacitly assumed throughout this paper that all of these Lorentz invariant products are 
in the Euclidean domain, i.e., $s_{jj+1}>0$, $m^2 > 0$. We will be interested in a particular set of indices in Eq.\ \re{PentaboxFI}, the one
that corresponds to an integral contributing to the five-leg Coulomb branch amplitude \cite{Belitsky:2024rwv}, namely $G_{11111111-100}$. 
This choice, although very particular, is not a limitation for the application of our analysis to other integrals in the family. We merely have to 
rerun our routines for other cases. Currently, we prefer to be specific rather than generic.

Introducing the Schwinger parameters $x_j$, the loop momenta can be integrated out, and we obtain the Feynman/Schwinger integral
\begin{align}
\label{FeynmanIntegralInput}
G_{11111111-100}
=
\left(\mu^{2}  \e^{\gamma_{\rm E}} \right)^{2 \ep}  \Gamma (3 + 2 \ep)
\int_{[0,\infty)} d^8 \bit{x}\;\delta\left(\sum_{j \in \mathcal{S}} x_j-1\right)\; \partial_{x_9} [ U^{1 + 3 \ep} F^{-3 - 2 \ep} ]|_{x_9 = 0}
\, ,
\end{align}
with the integration measure $d^8 \bit{x} = dx_1 \dots dx_8$ and the integrand determined by two Sy\-man\-zik polynomials, conventionally 
called $U$ and $F$. They are constructed as a sum of trees $T^1$ and two-trees $T^2$ (embellished with squares of momentum flows 
${\rm mom}^2_{T^2}$ from one tree component to another), respectively,
\begin{align}
\label{SymanzikPolys}
U(\bit{x}, x_9) = \sum_{T^1} \prod_{i \not\in T^1} x_i
\, , \qquad
F(\bit{x}, x_9; s_{jj+1}, m) = - \sum_{T^2} {\rm mom}^2_{T^2} \prod_{i \not\in T^2} x_i
\, .
\end{align}
We do not display them explicitly since they are very lengthy, consisting of 23 and 70 terms, respectively. But the reader can easily generate them
with the {\tt UF} routine of {\tt FIESTA} \cite{Smirnov:2021rhf}. The ${\rm GL}(1)$ invariance of the integration measure manifests itself in the Cheng-Wu 
theorem \cite{Cheng:1987ga}, i.e., a statement that a sum of any subset $\mathcal{S}$ of the Schwinger parameters can be set to 1. Practically, we 
will set just one of these equal to one.

\section{Near mass-shell limit with MofR}
\label{MofRsection}

The Feynman/Schwinger integral representation is the most suitable for a geometric approach to the analysis of the asymptotic behavior 
of the Feynman integrals $G$ as $m \to 0$ \cite{Smirnov:1999bza,Pak:2010pt}. Information about it is encoded in the Newton polytope
associated with the parametric integrand built out of the two Symanzik polynomials $U$ and $F$. For a concise analysis, it is sufficient to
consider their product rather than studying associated polytopes individually. We start by constructing the multivariate polynomial in Schwinger
parameters $x_j$ and the `small' invariant $m^2$
\begin{align}
\label{UFdef}
UF = \sum_j c_j m^{2 n_{0,j}} \bit{x}^{\bit{\scriptstyle n}_j} 
\, .
\end{align}
Here and below, we use the shorthand notation for the products of tuples of monomials
\begin{align}
\bit{x}^{\bit{\scriptstyle n}} \equiv x_1^{n_1} \dots x_8^{n_8}
\, .
\end{align}
The integer-valued $\bit{n}$-components are drawn from $\{0,1,2\}$, while $n_0$ is binary $\{0,1\}$. The set of `vertices' $(n_0, \bit{n})$
defines a hull. Due to the ${\rm GL} (1)$ invariance, we can set any $x_j$ to 1. 
In our particular case, without loss of generality, we can project, say, on the $n_8 = 0$ 
plane and define the Newton polytope as a convex hull of vertices
\begin{align}
{\rm Newt} [UF] = {\rm ConvHull} \{ (n_0, n_1, \dots, n_7) | n_0 \in \{ 0,1 \}, n_{1,\dots, 7} \in \{0,1,2\}  \}
\, .
\end{align}
All expansion coefficients $c_j$ accompanying the monomials in Eq.\ \re{UFdef} are positive definite due to the chosen Euclidean kinematics, 
so the above definition of the polytope is proper.

As $m \to 0$, the original loop-momentum integrals receive leading contributions from a particular set of momentum modes. In the 
current consideration, they are known as hard, collinear, and ultrasoft \cite{Belitsky:2024yag,Belitsky:2024rwv,Belitsky:2025bez}. 
These possess different scalings of their momenta in the `small' invariant $m$. Slicing the entire loop-momentum space into subspaces 
classified according to them defines the so-called {\sl regions} of MofR \cite{Beneke:1997zp}. Since Schwinger parameters 
are reciprocal to propagators, these momentum modes yield (reversely) equivalent scalings of $x_j$'s. In this manner, the Newton polytope 
of the integrand allows one to define the {\sl regions} in a completely geometrical fashion \cite{Pak:2010pt}.

The (outward-pointing) normals to the facets of the Newton polytope are the rays of its dual fan. The cones of the maximal 
dimension tile the entire seven-dimensional space. However, only a subset of these is relevant for the asymptotic behavior of the Feynman 
integral. Namely, the cones of the minimal dimension, i.e., the rays, that possess a negative projection\footnote{Notice that this criterion is
opposite to the one in Ref.\ \cite{Pak:2010pt} due to their use of inward-pointing normals. We use the opposite direction for a uniform 
treatment with other polytopes appearing later in our analysis.} on the 0-th direction define the leading {\rm regions}. Their components 
$\bit{r}$ projected on the $r_0 = 0$ plane provide the scaling law of corresponding Schwinger parameters\footnote{From now on, we will
set the dimensional regularization mass parameter to one, i.e., $\mu =1$. With this convention, the `small' and `large' invariants can
be regarded as dimensionless. In the total sum of all regions, the $\mu$-dependence will cancel, and the proper mass dimension of the 
Feynman integral restored.},
\begin{align}
\label{MofRSchwingerRescaling}
\bit{x} \to \big(\mu^2/m^2\big)^{\bit{\scriptstyle r}} \bit{x}
\, .
\end{align}

The above strategy is implemented in the {\tt asy} routine \cite{Pak:2010pt} that relies on the {\tt QHull} linear programming algorithm 
\cite{Barber:1996lmi} for finding the polytope's normals. It is part of {\tt FIESTA} and is called with {\tt SDExpandAsy} command.
It was employed by us to reveal 32 nontrivial\footnote{Trivial leading regions result in scaleless integrals which are set to zero within 
dimensional regularization.} leading regions. The Feynman integral \re{FeynmanIntegralInput} is then decomposed as
\begin{align}
\label{PentaboxRegionsRep}
G_{11111111-100}|_{m \to 0}
=
\sum_{i = 1}^{32} M_i {\rm Reg}_r + O (m)
\, ,
\end{align}
where the $m$-dependence is completely factored out from the accompanying parametric integrals into prefactors, given by the 32-dimensional 
array
\begin{align}
M
=
\big\{
&
m^{-8\ep},m^{-8\ep},m^0,m^{-2\ep},m^{-8\ep},m^{-8\ep},m^{-4\ep},m^{-6\ep},m^{-6\ep},m^{-4\ep},m^{-6\ep},m^{-6\ep},
\\
&
m^{-4\ep},m^{-4\ep},m^{-4\ep}, m^{-4\ep}, m^{-6\ep},m^{-6\ep},m^{-6 \ep},m^{-6\ep},m^{-4\ep},m^{-6\ep},m^{-2\ep},m^{-2\ep},
\nonumber\\
&\qquad\qquad\qquad\qquad\qquad\ \,
m^{-6 \ep},m^{-4\ep},m^{-6\ep},m^{-4\ep},m^{-4\ep},m^{-2\ep},m^{-2 \ep},m^{-4\ep}
\big\}
\, . \nonumber
\end{align}
These exponents are in one-to-one correspondence with the loop momentum scalings inducing various regions, i.e., $m^{- 2 (\ell_1 + \ell_2)}$
stem for $\ell_i = {0\, (\mbox{hard}),1\, (\mbox{collinear}), 2\, (\mbox{ultrasoft})}$ for one of two loop momenta. The reduced parametric 
integrals ${\rm Reg}_i$ develop poles in $\ep$ and are listed in Appendix \ref{RegionIntegrals}. Taylor expansion of $M$ in $\ep$ generates 
infrared logarithms in the particle's virtuality $m$.

\section{Tropical geometry of regions}
\label{TropicsSection}

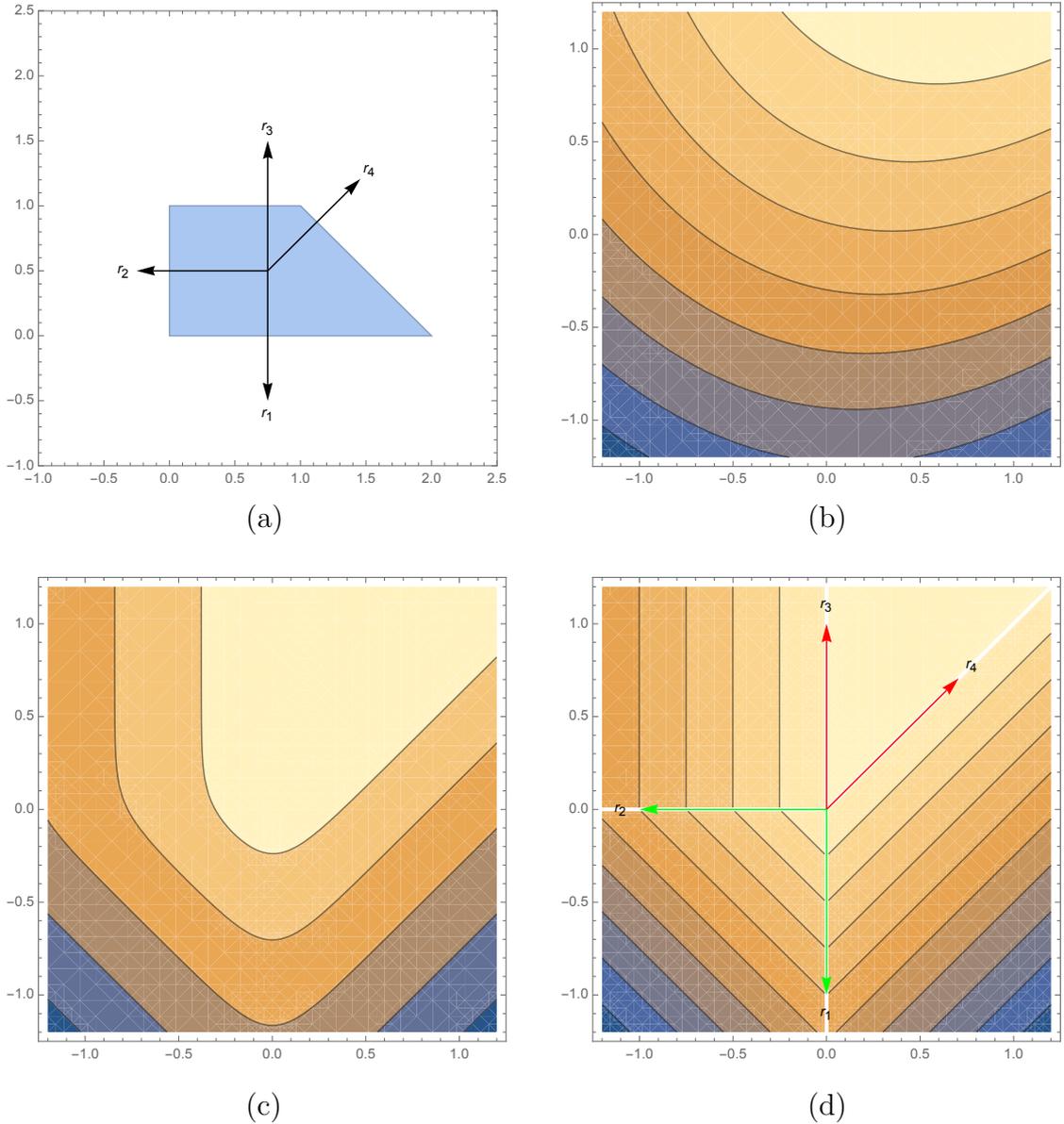
\begin{figure}[t]
\begin{center}
\mbox{
\begin{picture}(0,430)(220,0)
\put(0,240){\insertfig{7}{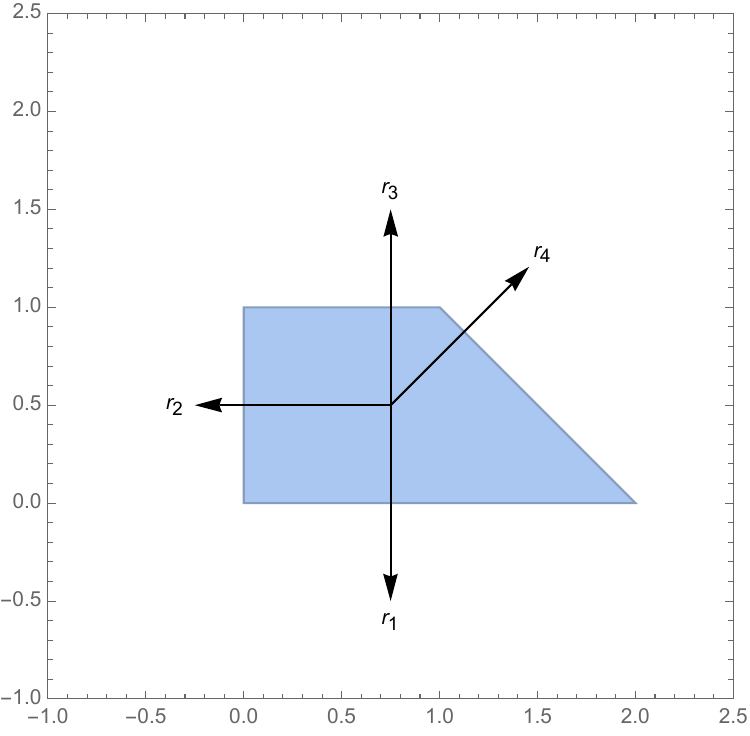}}
\put(220,240){\insertfig{7}{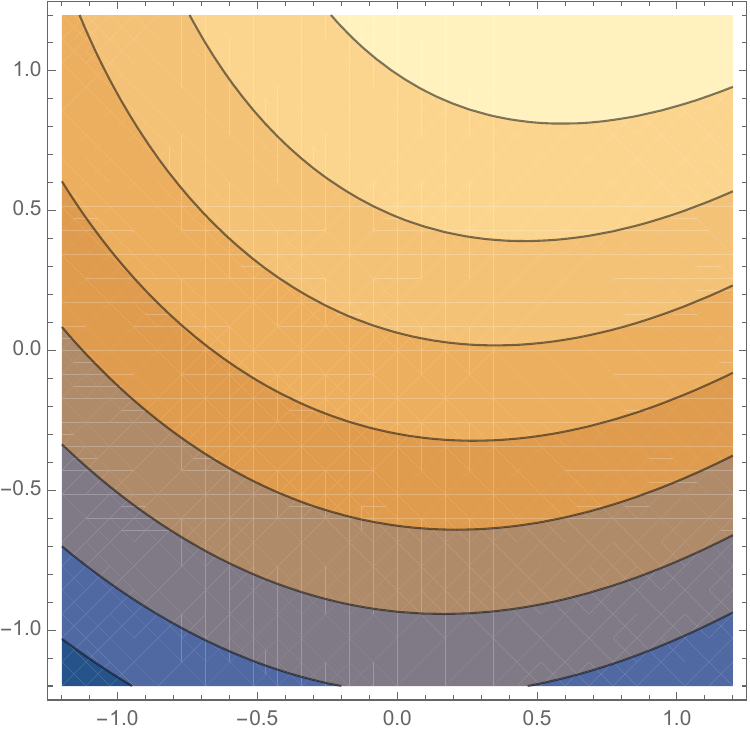}}
\put(0,10){\insertfig{7}{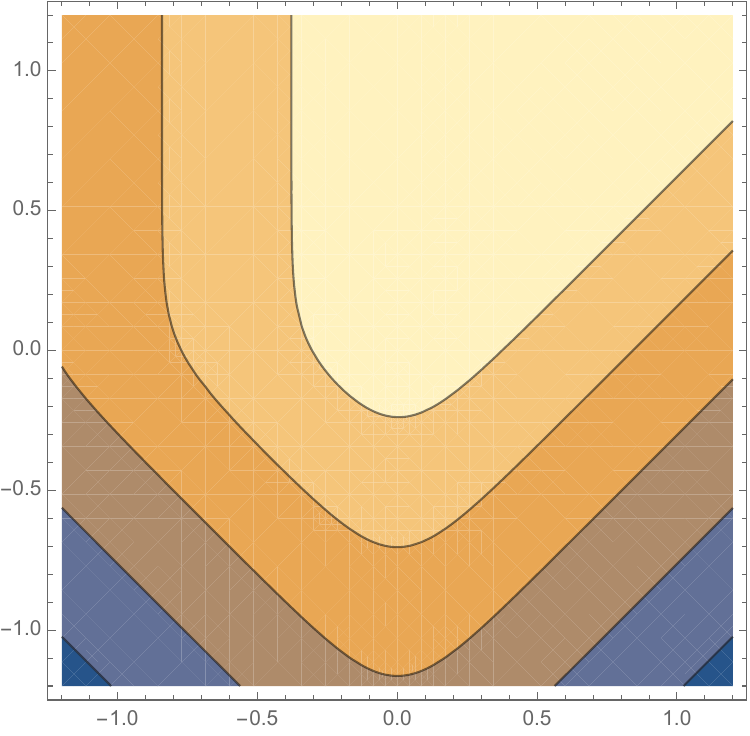}}
\put(220,10){\insertfig{7}{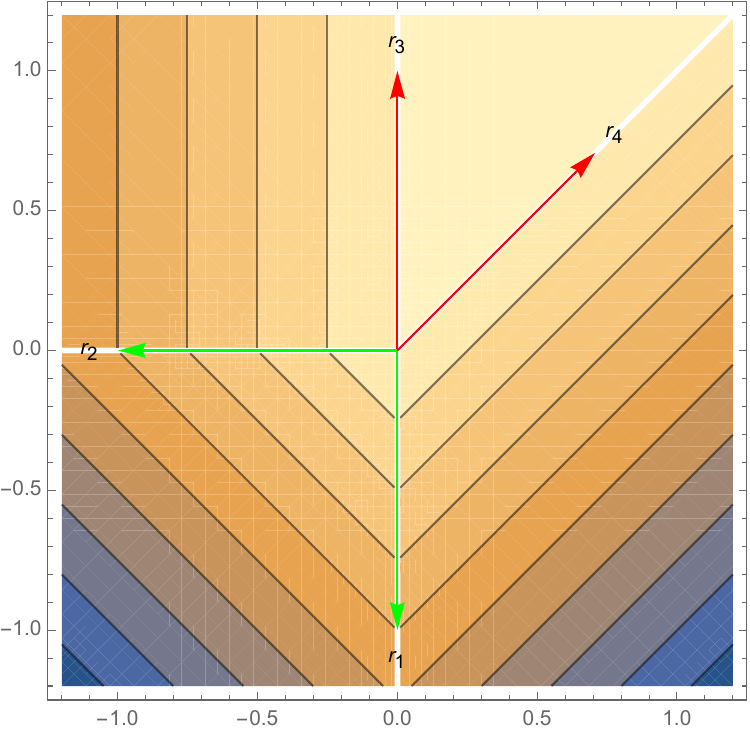}}
\put(95,225){(a)}
\put(318,225){(b)}
\put(95,-10){(c)}
\put(318,-10){(d)}
\end{picture}
}
\end{center}
\caption{\label{TropGeomReg18} The Newton polytope $\mathcal{P}_{18}$, see Eq.\ \re{P18}, with its facets' normals is in (a). Contour plots of the 
natural (b), and common (c) logarithmic maps $\bit{t} = \log_{\rm base} \bit{x}$ of the natural and common logarithms, respectively, of the 
integrand for region~18. The latter is larger/smaller in brighter yellow/deeper blue domains. As the base of the logarithm increases, the 
integrand begins to display sharper separation of its domains of linearity, becoming piece-wise in its tropical approximation in (d). The latter 
panel also shows the divergent (red) and regular (green) rays of the dual fan of the Newton polytope.}
\end{figure}

Having established the form of the reduced parametric integrals for all leading regions, we are now in a position to evaluate them. They are 
complicated multivariate functions of the Mandelstam-like invariants $s_{jj+1}$. However, we anticipate them to be given in terms of GPLs since 
a naked eye inspection suggests that the parametric integrands possess the property of linear reducibility, although it is not a given after 
performing a few successive integrations, as it can be easily broken along the way. We want to preserve it as long as we can. We will
see below what it entails. As we advocated in the introduction, this will take the sector decomposition off the table and will leave the 
Nilsson-Passare \cite{Nilsson,Berkesch} analytic continuation on the `back burner'. 
 
Let us provide a short introduction to the required elements of algebraic geometry, which form the foundation for the framework set up 
in Ref.\ \cite{Salvatori:2024nva}. The formalism is rooted in the unravelling of divergences through the understanding of the tropical geometry 
of Newton polytopes associated with given (reduced, in the current application) integrands. It furnishes singularities of Feynman integrals 
with a simple geometrical origin. The inception of the method appeared earlier in Ref.\ \cite{Arkani-Hamed:2022cqe}. 

The Euclidean kinematics chosen by us for all Lorentz invariants implies that our Feyn\-man/Schwinger parametric integrals can potentially 
diverge only at the boundaries of their semi-infinite integration domains; there are none in their bulk since all Symanzik polynomials are 
positive-definite. The former can be uncovered using the logarithmic map $\bit{t} = \log_{\rm base} \bit{x}$, which zooms in on the 
$\bit{t} \to \infty$ asymptotic behavior of integrands. Tropical arithmetic arises as a classical arithmetic under the logarithm when its 
base is taken to infinity.  It is known as Maslov's dequantization of positive real numbers, see Refs.\ 
\cite{Rau2017,2015arXiv150205950B,MikhalkinRau2019} for comprehensive profiles. 

Its relevance to the divergence structure of Feynman/Schwinger integrals can be visualized with an example. Take, for instance, 
the two-fold (after fixing the ${\rm GL} (1)$ gauge $x_5 = 1$) integral for region~18 \re{Reg18Init}. It will serve our demonstration 
purposes in the simplest, though non-trivial, circumstances. The main focus is on the region's integrand $\mathcal{I}_{18}$
\begin{align}
{\rm Reg}_{18} = c_{18} \int\frac{d^2 \bit{x}}{\bit{x}} \mathcal{I}_{18} (\bit{x})
\end{align}
accompanying the logarithmic measure in Schwinger parameters $d^2 \bit{x}/\bit{x} \equiv d \log x_1 d \log x_2$,
\begin{align}
\mathcal{I}_{18} (\bit{x}) = x_1 x_2^{1 - \ep} (1+x_1)^{-1-2 \ep} \frac{(1 + x_1 + x_2)^{2 \ep}}{(1 + x_1)/u + x_2}
\, .
\end{align}
The dependence on the external kinematics enters here through the ratio $u$ determined by the Mandelstam-like invariants $u = s_{12}/s_{45}$. 
Stripping the monomials $x_1 x_2^{1 - \ep}$ off the rest\footnote{Overall monomials are irrelevant for the Newton polytope's geometry since 
they merely translate the polytope in the ambient space without altering its shape.}, we can find the Newton polytope for this integrand as a 
Minkowski sum, see, e.g., Ref.\ \cite{z-lop-95,MartinEtAl1999,Oda1988} for a glossary of terms, of the convex hulls of vertices of the reduced 
Symanzik polynomials. For generic Mandelstam-like invariants, their value is irrelevant for the algebraic geometry analysis, which follows, 
so we can safely set them to unity. Thus
\begin{align}
\label{P18}
\mathcal{P}_{18}
\equiv
{\rm Newt} [\mathcal{I}_{18}]
&
= {\rm Newt} [1+x_1] \oplus {\rm Newt} [1 + x_1 + x_2]
\\
&
=
{\rm ConvHull} \{ \{2, 0\}, \{1, 1\}, \{1, 0\}, \{0, 1\}, \{0, 0\} \}
\, . \nonumber
\end{align}
This Newton polytope is shown in the panel (a) of Fig.\ \ref{TropGeomReg18} along with a set of normals $\{ \bit{r}_1, \dots, \bit{r}_4\}$ to the 
polytope's (here, polygon's) facets (here, edges). For the logarithmic map $\bit{t} = \log_{\rm base} \bit{x}$ of $\log_{\rm base} 
\mathcal{I}_{18} (\bit{x})$, the domains of its linearity become more and more prominent and well separated, as we increase the base. 
The limit as ${\rm base} \to \infty$ defines its {\sl tropical} approximation,
\begin{align}
\label{TropIntegrand}
\mbox{Trop} [\mathcal{I}_{18}]
&
=
\lim_{{\rm base} \to \infty}
\log_{\rm base} \mathcal{I}_{18} |_{\bit{\scriptstyle x} = {\rm base}^{\bit{\scriptstyle t}}}
\\
&
= t_1 + (1 - \ep) t_2 - (1 + 2 \ep) \max (0, t_1) - (1 - 2 \ep) \max (0, t_1, t_2)
\nonumber
\, .
\end{align}
We will simply refer to it as $\mbox{Trop} (\bit{t}) = \mbox{Trop} [\mathcal{I}_{18}]$ for the rest of this section.

The normals to the polytope's facets define the boundaries of the aforementioned domains of the integrand's linearity. These normals
are the one-dimensional cones, aka rays, of the dual fan of the Newton polytope \cite{z-lop-95,Oda1988}. Choosing them on an integer
lattice as close as possible to the origin, they read explicitly for region 18
\begin{align}
\label{RaysReg18}
\bit{r} = \{\{0, -1\}, \{-1, 0\}, \{0, 1\}, \{1, 1\}\}
\, .
\end{align}
These play a pivotal role in unravelling the divergent structure of the Feynman/Schwinger integral. The cones of highest dimension, 
i.e., two, in the current case, are {\sl generated} by these rays and tile the entire integration domain 
\begin{align}
\label{Triangulation}
\mathbb{R}^2 = {\rm Span}_+ (\bit{r}_1, \bit{r}_2) \cup {\rm Span}_+ (\bit{r}_2, \bit{r}_3) 
\cup 
{\rm Span}_+ (\bit{r}_3, \bit{r}_4) \cup {\rm Span}_+ (\bit{r}_4, \bit{r}_1)
\, .
\end{align}
This defines a {\sl triangulation}.
It is enough, however, to ascertain the behavior of the integrand, through its tropical approximation \re{TropIntegrand}, only on the rays $\bit{r}$, 
rather than keeping track of it at every point of $\mathbb{R}^2$ \cite{Arkani-Hamed:2022cqe,Salvatori:2024nva}. Evaluating \re{TropIntegrand} 
on \re{RaysReg18}, we find
\begin{align}
\mbox{Trop} (\bit{r}) = \{ - 1 + \ep, -1, \ep, - \ep\}
\, .
\end{align}
In this manner, we associate a number with each facet (dual to a given ray) of the polytope that determines its degree of divergence 
\cite{Arkani-Hamed:2022cqe}. Recalling that this behavior of the logarithm of the integrand at large values of Schwinger parameters is 
accompanied by the logarithmic measure, the first two values with ${\rm Trop} < 0$ yield convergent behavior of the integral, while the 
latter two generate logarithmic divergences for $\ep = 0$, ${\rm Trop} = 0$. ${\rm Trop} > 0$ will not appear in our study. This consideration 
suggests that the original integral is divergent on the subfan $\Sigma_{\rm div}$ of the original dual fan of the Newton polytope
\begin{align}
\label{DivSubFan}
\Sigma_{\rm div}
=
\{ {\rm Span}_+ (\bit{r}_3), {\rm Span}_+ (\bit{r}_4), {\rm Span}_+ (\bit{r}_3, \bit{r}_4) \}
\, .
\end{align}
This subfan is formed by {\sl simplicial} cones since their {\sl generators} $\bit{r}_{3,4}$ are linearly independent. This property will be 
of paramount importance in the next section.

The tropical approximation of the integrand captures the asymptotic behavior of the latter for large values of Schwinger parameters and 
relates it to geometrical data of the Newton polytope, i.e., its divergent fan \re{DivSubFan}. To determine the value of the integral more 
accurately, this information is not sufficient, however. We need the coefficients accompanying this scaling. These are determined by the 
monomials in the integrand's Symanzik polynomials that stay piece-wise constant in the divergent subfan. These monomials are the ones 
with the exponent vectors $\bit{n}$ that belong to the corresponding face of the Newton polytope. They are determined by the {\sl initial forms}. 
For a ray, it is evaluated as
\begin{align}
\label{IFray}
\mathcal{I}_{18}|_{\bit{\scriptstyle r}} 
= \lim_{L \to \infty} 
L^{- {\rm Trop}(\bit{\scriptstyle r})} \mathcal{I}_{18} \left(L^{\bit{\scriptstyle r}}  \bit{x} \right) 
\, ,
\end{align} 
while for a two-dimensional cone ${\rm Span}_+ (\bit{r}_j, \bit{r}_k)$, it is
\begin{align}
\label{IF2dCone}
\mathcal{I}_{18}|_{ {\rm Span}_+ (\bit{\scriptstyle r}_j, \bit{\scriptstyle r}_k)} 
=
\lim_{L_1 \to \infty} 
L_1^{- {\rm Trop} (\bit{\scriptstyle r}_k )}
\lim_{L_2 \to \infty} 
L_2^{- {\rm Trop}(\bit{\scriptstyle r}_j ) }
\mathcal{I}_{18} \left( L_1^{\bit{\scriptstyle r}_k} L_2^{\bit{\scriptstyle r}_j} \bit{x} \right) 
\, .
\end{align}
For higher-dimensional cones, when present, the generalization is straightforward. For compatible rays of the fan, the order of taking 
the limits is irrelevant. Rays' compatibility can be read off from the lattice structure of the fan, through its Hasse graph (to be 
demonstrated later on). 

The cones of the maximal dimensions in the divergent fan determine the leading singularity of the Feynman/Schwinger parametric 
integral in question, as was already employed in Ref.\ \cite{Arkani-Hamed:2022cqe}. Introducing the barycentric coordinates, see, 
e.g., \cite{Kaneko:2009qx}, which read in terms of the original coordinates
\begin{align}
\label{BarycentricVars}
\log \bit{x} = - \bit{T} \cdot \log \bit{y}
\, , 
\end{align}
with the unimodular transformation matrix $\bit{T} = (\bit{r}_3|\bit{r_4})$ formed by the divergent rays listed as columns, the $\bit{y}$-integrations 
can be easily performed with the result
\begin{align}
\label{LeadingSingularityReg18}
{\rm Reg}_{18}|_{\rm leading \ singularity} 
&= 
c_{18} \int_{{\rm Span}_+ (\bit{\scriptstyle r}_3, \bit{\scriptstyle r}_4)}
\frac{d^2 \bit{x}}{\bit{x}} \mathcal{I}_{18}|_{{\rm Span}_+ (\bit{\scriptstyle r}_3, \bit{\scriptstyle r}_4)} (\bit{x})
\nonumber\\
&
=
\frac{c_{18}}{[-{\rm Trop} (\bit{r}_3)][-{\rm Trop} (\bit{r}_4)]}
\, .
\end{align}
Attempts to extend this consideration to other cones in the fan inevitably lead to the violation of linear reducibility, a property 
we would like to preserve.

\section{Locally finite integrands}
\label{LocFiniteSection}

The analysis in the previous section is as far as the tropical geometry alone can take us. The triangulation of the integration domain with 
subsequent barycentric change of variables is well-suited for numerical algorithms \cite{Kaneko:2009qx,Smirnov:2021rhf,Borowka:2017idc}.
However, this route is doomed for analytical approaches, which aim at a systematic extraction of poles with subsequent determination of 
accompanying coefficients and finite remainders as functions of kinematical variables.

This calls for the abandonment of the space triangulation. However, integrals over the entire integration domain with integrands 
approximated in a given cone of the dual fan will be singular regardless of the initially implemented dimensional regularization. 
Thus, one has to either impose an extra regularization when splitting the integrand according to its tropical properties, or `subtract'
these divergences at the integrand level in such a manner that their inclusion/exclusion to/from the original integral yields well-defined 
individual contributions. The second route is taken in Ref.\ \cite{Salvatori:2024nva} and leads to the notion of {\sl locally finite} 
integrands. The latter implies that the expansion in $\ep$ of the integral can be obtained by the expansion of the corresponding 
integrand. The procedure of representing a given integral as a linear combination of integrals with locally finite integrands is based 
on some `subtractions' which differ in character from the conventionally-understood subtraction of poles in $\ep$ within dimensional 
renormalization. In our procedure, these poles are not removed; rather, they are merely extracted from integrals to arrive at locally finite 
integrands\footnote{This is the reason for us to use the notions such as `subtractions' and ‘counterterms’ in quotes to avoid confusion with
traditional terminology.}.
 
To motivate how this has to be enabled, let us focus on the leading singularity \re{LeadingSingularityReg18}. If the integration domain in that
integral is enlarged from ${\rm Span}_+ (\bit{r}_3, \bit{r}_4)$ to the entire quadrant $0 \leq \bit{x} < \infty$, the integral will be singular on the lower 
limits since the initial form of the integrand captures correctly its large $\bit{x} \to \infty$ asymptotics only: it is oblivious to the behavior in all 
other cones of the dual fan \re{Triangulation}. How do we make it well-defined in the entire domain? We can do it with the introduction of a 
`suppression' factor. Since we do not want to alter the large-$\bit{x}$ asymptotics of the integrand, it has to go to one there, while 
simultaneously this prefactor has to provide sufficient suppression in the singular vicinity of the origin to furnish the integral with a finite 
regularized value\footnote{To put this into perspective, recall that a similar suppression was introduced within the context of the
Bogolyubov-Parassyuk-Hepp-Zimmermann ultraviolet renormalization \cite{Bogoliubov:1957gp,Hepp:1966eg,Zimmermann:1969jj} based 
on integrand subtractions at zero external momenta. A modification of this scheme employing auxiliary masses to regulate emerging infrared 
divergences was also proposed in the literature \cite{Lowenstein:1975pd}.}. Of 
course, there is a plethora of choices that fit the profile. However, this is not the most stringent requirement on its form. Since the term we are 
currently discussing extracts the leading singularity from the Feynman/Schwinger integral, it has to be `subtracted' from the original expression 
to determine the remainder. We want this subtraction (i) not to produce unregularized integrals and only (ii) generate less singular terms in the 
parameter of dimensional regularization. Moreover, we have to (iii) preserve the property of linear reducibility to stand a chance of calculating 
the rest analytically. This cumulative information implies that the sought-after `suppression' factor should be a rational function of Schwinger 
parameters, and its properties have to inherit those of the divergent subfan.

The form of the `suppression' factor that obeys the above properties is not unique, but an ansatz was proposed in Ref.\ \cite{Salvatori:2024nva}.
Echoing the tropicalization of cross-ratio variables, which appeared in a related context of ultraviolet divergences \cite{Hillman:2023ezp,Hillman:2023vas},
as well as string amplitudes \cite{Brown:2019wna} and stringy canonical forms \cite{Arkani-Hamed:2019mrd}, we define a set of `conjugate' 
vectors $\bit{w}_j$ obeying the so-called {\sl geometric} property \cite{Salvatori:2024nva}
\begin{align}
\label{GeometricProperty}
\bit{r}_j \cdot \bit{w}_k = - \delta_{jk}
\, ,
\end{align}
for all rays $\bit{r}$,  in the divergent subfan $\Sigma_{\rm div}$, and introduce the `suppression' variables as
\begin{align}
\label{vVariables}
v_{ \bit{\scriptstyle r}_j} = (1 + \bit{x}^{\bit{\scriptstyle w}_j})^{-1}
\, .
\end{align}
Then, a `counterterm' stemming from a cone ${\rm Span}_+ (\bit{r}_j, \bit{r}_k, \dots)$ is \cite{Salvatori:2024nva}
\begin{align}
\mathcal{I}|_{{\rm Span}_+ (\bit{\scriptstyle r}_j, \bit{\scriptstyle r}_k, \dots)}
\, 
v_{ \bit{\scriptstyle r}_j }^{-{\rm Trop} (\bit{\scriptstyle r}_j)}
\, 
v_{ \bit{\scriptstyle r}_j }^{-{\rm Trop} (\bit{\scriptstyle r}_j)}
\dots 
\, .
\end{align}
The existence of a basis of $\bit{w}$'s implies that the divergent subfan has to be {\sl simplicial}. When this is not the case, 
further preparatory steps are required. These will be addressed in the next sections.

To conclude our analysis of the contribution from region 18, we are now ready to calculate it as a Laurent series in $\ep$ up to $O(\ep)$ and 
determine its coefficient analytically. The inclusion/exclusion allows us to write down the initial expression in terms of a sum
\begin{align}
\label{FiniteIntsExp18}
{\rm Reg}_{18} 
&
= c_{18} \int \frac{d^2 \bit{x}}{\bit{x}}
\big[
\mathcal{I}^{(1)}_{18} + \mathcal{I}^{(2)}_{18} + \mathcal{I}^{(3)}_{18} + \mathcal{I}^{(4)}_{18}
\big]
\, ,
\end{align} 
of the following integrands
\begin{align}
\mathcal{I}^{(1)}_{18}
&
=
\mathcal{I}_{18} 
- 
v_3^{-{\rm Trop}_3} \mathcal{I}_{18} |_{\bit{\scriptstyle r}_3} 
- 
v_4^{-{\rm Trop}_4} \mathcal{I}_{18} |_{\bit{\scriptstyle r}_4}
+
v_3^{-{\rm Trop}_3}
v_4^{-{\rm Trop}_4}
\mathcal{I}_{18}|_{{\rm Span}_+ (\bit{\scriptstyle r}_3, \bit{\scriptstyle r}_4)}
\, ,
\nonumber\\
\mathcal{I}^{(2)}_{18}
&
=
v_3^{-{\rm Trop}_3} 
\big[  
\mathcal{I}_{18} |_{\bit{\scriptstyle r}_3} 
-
v_4^{-{\rm Trop}_4}
\mathcal{I}_{18}|_{{\rm Span}_+ (\bit{\scriptstyle r}_3, \bit{\scriptstyle r}_4)}
\big]
\, ,
\nonumber\\
\mathcal{I}^{(3)}_{18}
&
=
v_4^{-{\rm Trop}_4}
\big[ 
\mathcal{I}_{18} |_{\bit{\scriptstyle r}_4}
-
v_3^{-{\rm Trop}_3}
\mathcal{I}_{18}|_{{\rm Span}_+ (\bit{\scriptstyle r}_3, \bit{\scriptstyle r}_4)}
\big]
\, ,
\nonumber\\
\mathcal{I}^{(4)}_{18}
&
=
v_3^{-{\rm Trop}_3}
v_4^{-{\rm Trop}_4}
\mathcal{I}_{18}|_{{\rm Span}_+ (\bit{\scriptstyle r}_3, \bit{\scriptstyle r}_4)}
\, ,
\end{align}
where, to simplify the presentation, we used only the rays' indices to label corresponding variables rather than displaying the former as their
arguments. Here, the initial forms, calculated according to Eqs.\ \re{IFray} and \re{IF2dCone}, read
\begin{align}
&
\mathcal{I}_{18} |_{\bit{\scriptstyle r}_3}
=
x_1 (1 + x_1)^{-1 - 2 \ep} x_2^{\ep}
\, , \\
&
\mathcal{I}_{18} |_{\bit{\scriptstyle r}_4}
=
x_1^{-2 \ep} (x_1 + x_2)^{2 \ep} x_2^{1 - \ep}
/(
x_1/u + x_2
)
\, , \\
&
\mathcal{I}_{18}|_{{\rm Span}_+ (\bit{\scriptstyle r}_3, \bit{\scriptstyle r}_4)}
=
x_1^{- 2 \ep} x_2^{\ep}
\, .
\end{align}
The first integrand $\mathcal{I}^{(1)}_{18}$ is locally finite. The rest are `counterterms'. But after the extraction of poles from 
these, their remainder will be finite as well. Let us demonstrate this explicitly since this region is simple enough to fit everything on
one page. 

Recalling that 
\begin{align}
c_{18} = \frac{ \e^{2 \ep \gamma_{\rm E}} s_{34}^\ep}{s_{12} s_{23} s_{45}} \Gamma (\ep) \Gamma (1-\ep) \Gamma(2 \ep)
\, ,
\end{align}
we need to evaluate all integrals to order $O (\ep^3)$. We start with the integrand $\mathcal{I}^{(1)}_{18}$, which is locally finite,
so we can immediately expand it in the Taylor series in $\ep$ and feed it into {\tt HyperInt}  \cite{Panzer:2014gra}. The latter calculates 
it without a hiccup and yields
\begin{align}
\int \frac{d^2 \bit{x}}{\bit{x}}
\mathcal{I}^{(1)}_{18}
=
- \ft16 \pi ^2 + 2 \zeta_3  \ep - \ft{1}{40} \pi^4 \ep^2 + O(\ep^3)
\, .
\end{align}
In this particular situation, one can get an $\ep$-exact result by noticing that the initial form $\mathcal{I}_{18} |_{\bit{\scriptstyle r}_4}$
entirely `subtracts' out the $u$-dependence of the original integrand, such that the resulting integral is $u$-independent and 
can be calculated in a closed form
\begin{align}
\int \frac{d^2 \bit{x}}{\bit{x}}
\mathcal{I}^{(1)}_{18}
=
\frac{\Gamma^2(\ep)}{\Gamma(1 + 2 \ep)} - \frac{1}{\ep^2}
\, .
\end{align}

The second term in Eq.\ \re{FiniteIntsExp18} is the first `counterterm'. So, it develops a pole in $\ep$. The latter can be easily extracted 
from it, making use of the barycentric variable transformation \re{BarycentricVars} with the unimodular matrix
\begin{align}
\bit{T} = (\bit{r}_3| \bit{e}_1)
\, , 
\end{align}
where we complemented the divergent ray $\bit{r}_3$ with a unit vector $\bit{e}_1 = (1,0)$. Then the $y_1$ integration can be 
performed in a straightforward manner, producing an overall factor of the regularized `volume' of the divergent ray, 
$\mbox{Vol}(\bit{r}_3) = 1/[- \mbox{Trop}_3] = - 1/\ep$, with the remaining integrand being locally finite. In fact, it vanishes 
for $\ep = 0$, so it is $o(\ep)$. We find
\begin{align}
\int \frac{d^2 \bit{x}}{\bit{x}}
\mathcal{I}^{(2)}_{18}
=
\frac{1}{[-\mbox{Trop}_3]} \int_0^\infty dy_2 \, y_2^{-1 + \ep} 
\left[
(1 + y_2)^{-1 - 2 \ep}
-
(1 + y_2)^{-1 - \ep}
\right]
=
\frac{1}{\ep^2}
- 
\frac{\Gamma^2 (\ep)}{\Gamma (1 + 2\ep)}
\, .
\end{align}
The integral here was simple enough to be evaluated exactly in $\ep$. Notice that the sum of the integrated $\mathcal{I}^{(1)}_{18}$ 
and $\mathcal{I}^{(2)}_{18}$ vanishes, which implies that the inclusion/exclusion sometimes `oversubtracts'. However, though it requires
extra integrations to be performed, the procedure never fails.

The third term is treated in the very same fashion. First, we extract the regularized `volume' of the ray $\bit{r}_4$, $\mbox{Vol}(\bit{r}_4) = 
1/[- \mbox{Trop}_4]$, with the transformation matrix $\bit{T} = (\bit{r}_4| \bit{e}_1)$. Then, the remaining integral is again finite, and its integrand 
can be safely expanded in $\ep$. Its individual terms are more complex, though, since they now depend on the kinematical invariant $u$, but 
{\tt HyperInt} comes to the rescue and produces with ease 
\begin{align}
\int \frac{d^2 \bit{x}}{\bit{x}}
\mathcal{I}^{(3)}_{18}
&
=
\frac{1}{[- \mbox{Trop}_4]}
\Big[
\log u
+
\ep 
\Big(
\ft16 \pi^2 - \ft12 \log^2 r - \mbox{Li}_2 (1 - u)
\Big)
\\
+
\ep^2
\Big(
4 \zeta_3
&
+
\ft{1}{6} \pi^2 \log u - \log^2 u \log (1-u) + \ft{1}{6} \log^3 u - 2 \mbox{Li}_3 (u) - 4 \mbox{Li}_3 (1-u)
\Big)
+
\dots
\Big]
\, . \nonumber
\end{align}
We dropped the $o(\ep^3)$ term since it requires more than a couple of lines to be displayed, but does not bring any new insights. {\tt HyperInt} 
can handle arbitrarily high orders in $\ep$. Above, we also employed the package {\tt gtolrules.m} \cite{Frellesvig:2016ske} to convert all GPLs to 
classical polylogarithms. The full expression can be found in the accompanying notebook in terms of GPLs. 

Last but not least, the final `counterterm' is
\begin{align}
\int \frac{d^2 \bit{x}}{\bit{x}}
\mathcal{I}^{(4)}_{18}
=
\frac{1}{[-\mbox{Trop}_3][-\mbox{Trop}_4]}
\, .
\end{align}
It coincides with Eq.\ \re{LeadingSingularityReg18} defining the leading singularity. Summing all contributions together, we can verify the
correctness of our result by a numerical test with {\tt FIESTA}.

\section{Nonsimplicial divergent fan}
\label{NonsimplicialFanSection}

The above example, although, shows all of the ingredients of the formalism, is not representative enough to exhibit potential pitfalls. This 
is what we are turning to next. We, therefore, continue with a situation where the geometric property \re{GeometricProperty} is violated and 
how to go about it. As was alluded to earlier, this happens when the divergent subfan is not simplicial. As a rule of thumb, this problem
will be encountered more often than not when the number $E$ of independent Schwinger parameters in the Feynman/Schwinger 
integrals is of order or larger than twice the perturbative order $L$. This reflects the apparent\footnote{We will dwell on this notion below.} 
complexity of the integral in question. To get a quantitative estimate, let us assume that the divergent contribution of a given region 
saturates the maximal order of the pole in $\ep$. It is $2L$ at $L$ loops. Then, if a Feynman/Schwinger integral is accompanied by Euler 
gammas with $\ell$-poles in this region, these need to be subtracted from $2L$. The remaining poles should come from integrations 
over $E$ Schwinger parameters. This $E$ sets the dimension of the ambient space. The leading singularity of the integral comes from the 
volume of a maximal cone in the divergent fan formed by {\sl compatible} divergent rays. Consistency requires its order to be $2L - \ell$.
However, the total number $d$ of divergent rays in a fan can, of course, be greater than $d \geq 2L -\ell$ since a bulk of divergent rays
can be incompatible with respect to the fan. For the divergent fan to be simplicial, the rays have to be linearly independent, i.e., $d \leq E$
is a necessary condition. For a very rough estimate of $d$, recall that the total number of rays in a dual fan scales as $2^E$ for 
Feynman/Schwinger integrals being generalized permutahedra \cite{Schultka:2018nrs}, up to a stipulation that region integrals are 
generally not permutahedra due to their exceptional kinematics. The latter implies a reduction in the number of vertices in the polytope 
and, therefore, facets. Thus, the above is an overestimation. Taking $d \lesssim 2^{E}$ as an upper limit leads to the immediate violation 
of the geometric property as a consequence of the very fast growth of the number of divergent rays compared to the dimension of the 
ambient space.

\subsection{Tropical geometry of region 7}

Let us consider a specific, simple, yet more advanced example where the above situation occurs. We will intentionally overcomplicate 
it to demonstrate emerging difficulties. Consider region 7 as it comes out of {\tt asy}, see Eq.\ \re{Reg7Init}. The dimension of this 
integral can be significantly reduced, making use of the substitutions $x_6 = \eta \xi$ and $x_7 = \eta (1 - \xi)$ and integrating both $\eta$ 
and $\xi$ out. Further fixing the ${\rm GL} (1)$ redundancy, we would be left with a one-fold integral. We will not do this, however. 
Instead, we will merely fix the gauge $x_2 = 1$ and then apply the tropical geometry to the resulting integrand of apparently high complexity.

The starting point is then
\begin{align}
{\rm Reg}_{7} = c_{7} \int\frac{d^3 \bit{x}}{\bit{x}} \mathcal{I}_{7} (\bit{x})
\end{align}
with the integrand\footnote{We preserve the labelling of Schwinger parameters according to the propagators \re{Propragators} of the original 
Feynman momentum integral \re{PentaboxFI}.}
\begin{align}
\label{Region7Integrand}
\mathcal{I}_7 (\bit{x})
=
(1 + x_1)^{3 \ep} (x_6+x_7)^{3 \ep} (x_6 x_7 + x_1 x_6 + x_1 x_7 + x_1 x_6 x_7)^{-2 \ep}/(1 + x_1/u)
\, .
\end{align}
The parameter $u$ is the very same ratio that we encountered in the previous section, $u = s_{12}/s_{45}$. The overall coefficient
\begin{align}
c_7
=
\frac{\e^{2 \gamma_{\rm E}  \ep} \Gamma (2 \ep)}{s_{12} s_{23} s_{45}}
\end{align}
already possesses a pole in $\ep$ such that the anticipated other three should stem from the three-fold integral with the rescaling-invariant 
logarithmic measure
\begin{align}
\label{Measure}
\frac{d^3 \bit{x}}{\bit{x}} \equiv \frac{d x_1}{x_1}  \frac{d x_6}{x_6}  \frac{d x_7}{x_7}
\, .
\end{align}
From our discussion given in the preamble to this section, the (overestimated) upper limit on the number of rays in the fan for this integrand 
is $8=2^3$. So, if the number of divergent rays in it is anywhere near $8$, then we have a tension with the geometric property, which requires 
their number to be no more than 3.		

\begin{figure}[t]
\begin{center}
\mbox{
\begin{picture}(0,255)(120,0)
\put(0,0){\insertfig{8}{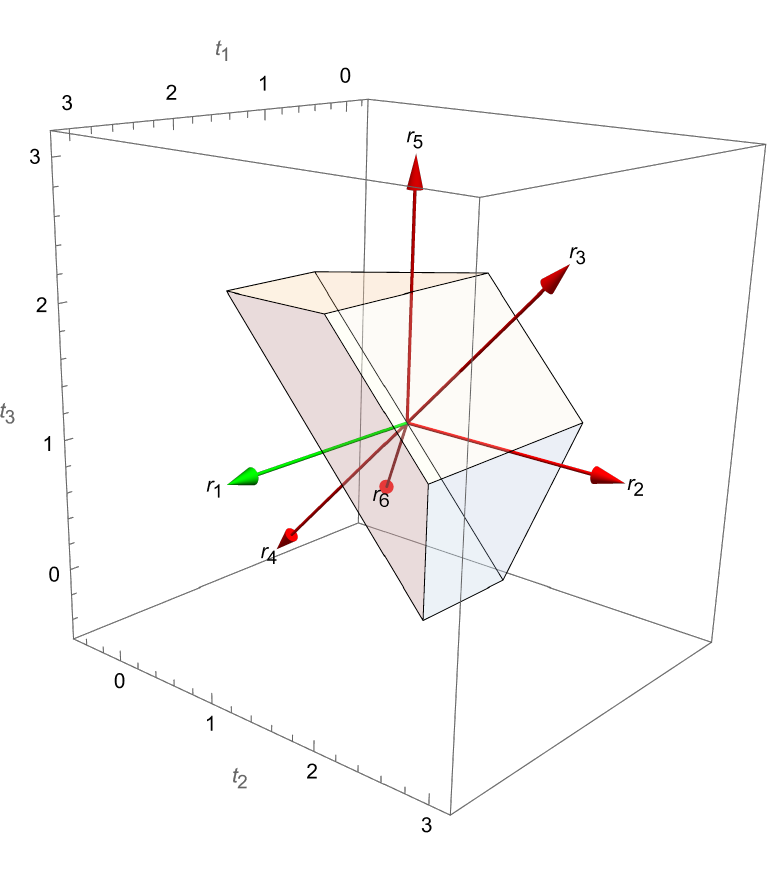}}
\end{picture}
}
\end{center}
\caption{\label{NewtonPolytope} The Newton polytope $\mathcal{P}_7$ \re{P7} for the region's 7 integrand \re{Region7Integrand} and 
corresponding normals to facets. Divergent and regular rays are shown in red and green, respectively. They form one-dimensional cones 
of the dual fan.}
\end{figure}

\begin{figure}[t]
\begin{center}
\resizebox{0.4\textwidth}{!}{
\begin{tikzpicture}
\put(-220,-130){

\begin{tikzpicture}[x  = {(1em, 0em)},
                    y  = {(0em, 10em)},
                    scale = 1,
                    color = {lightgray}]

  \coordinate (v0_unnamed__1) at (0, 3);
  \coordinate (v1_unnamed__1) at (-12, 2);
  \coordinate (v2_unnamed__1) at (-2.4, 2);
  \coordinate (v3_unnamed__1) at (-16.8, 2);
  \coordinate (v4_unnamed__1) at (-7.2, 2);
  \coordinate (v5_unnamed__1) at (16.8, 2);
  \coordinate (v6_unnamed__1) at (2.4, 2);
  \coordinate (v7_unnamed__1) at (7.2, 2);
  \coordinate (v8_unnamed__1) at (12, 2);
  \coordinate (v9_unnamed__1) at (-11.2, 1);
  \coordinate (v10_unnamed__1) at (-17.6, 1);
  \coordinate (v11_unnamed__1) at (-1.6, 1);
  \coordinate (v12_unnamed__1) at (-4.8, 1);
  \coordinate (v13_unnamed__1) at (8, 1);
  \coordinate (v14_unnamed__1) at (-14.4, 1);
  \coordinate (v15_unnamed__1) at (-8, 1);
  \coordinate (v16_unnamed__1) at (1.6, 1);
  \coordinate (v17_unnamed__1) at (14.4, 1);
  \coordinate (v18_unnamed__1) at (17.6, 1);
  \coordinate (v19_unnamed__1) at (4.8, 1);
  \coordinate (v20_unnamed__1) at (11.2, 1);
  \coordinate (v21_unnamed__1) at (3.10204, 0);
  \coordinate (v22_unnamed__1) at (-11.7551, 0);
  \coordinate (v23_unnamed__1) at (-3.91837, 0);
  \coordinate (v24_unnamed__1) at (4.70204, 0);
  \coordinate (v25_unnamed__1) at (-2.31837, 0);
  \coordinate (v26_unnamed__1) at (11.7551, 0);
  \coordinate (v27_unnamed__1) at (0, -1);

  \definecolor{vertexcolor_unnamed__1_0}{rgb}{ 0 0 0 }
  \definecolor{vertexcolor_unnamed__1_1}{rgb}{ 1 1 1 }

  \definecolor{vertexbordercolor_unnamed__1}{rgb}{ 0 0 0 }

  \tikzstyle{vertexstyle_unnamed__1_0} = [text=black, inner sep=2pt, rectangle, rounded corners=3pt,fill=vertexcolor_unnamed__1_0, draw=vertexbordercolor_unnamed__1,]
  \tikzstyle{vertexstyle_unnamed__1_1} = [text=black, inner sep=2pt, rectangle, rounded corners=3pt,fill=vertexcolor_unnamed__1_1, draw=vertexbordercolor_unnamed__1,]
  \tikzstyle{vertexstyle_unnamed__1_2} = [text=black, inner sep=2pt, rectangle, rounded corners=3pt,fill=vertexcolor_unnamed__1_1, draw=vertexbordercolor_unnamed__1,]
  \tikzstyle{vertexstyle_unnamed__1_3} = [text=black, inner sep=2pt, rectangle, rounded corners=3pt,fill=vertexcolor_unnamed__1_1, draw=vertexbordercolor_unnamed__1,]
  \tikzstyle{vertexstyle_unnamed__1_4} = [text=black, inner sep=2pt, rectangle, rounded corners=3pt,fill=vertexcolor_unnamed__1_1, draw=vertexbordercolor_unnamed__1,]
  \tikzstyle{vertexstyle_unnamed__1_5} = [text=black, inner sep=2pt, rectangle, rounded corners=3pt,fill=vertexcolor_unnamed__1_1, draw=vertexbordercolor_unnamed__1,]
  \tikzstyle{vertexstyle_unnamed__1_6} = [text=black, inner sep=2pt, rectangle, rounded corners=3pt,fill=vertexcolor_unnamed__1_1, draw=vertexbordercolor_unnamed__1,]
  \tikzstyle{vertexstyle_unnamed__1_7} = [text=black, inner sep=2pt, rectangle, rounded corners=3pt,fill=vertexcolor_unnamed__1_1, draw=vertexbordercolor_unnamed__1,]
  \tikzstyle{vertexstyle_unnamed__1_8} = [text=black, inner sep=2pt, rectangle, rounded corners=3pt,fill=vertexcolor_unnamed__1_1, draw=vertexbordercolor_unnamed__1,]
  \tikzstyle{vertexstyle_unnamed__1_9} = [text=black, inner sep=2pt, rectangle, rounded corners=3pt,fill=vertexcolor_unnamed__1_1, draw=vertexbordercolor_unnamed__1,]
  \tikzstyle{vertexstyle_unnamed__1_10} = [text=black, inner sep=2pt, rectangle, rounded corners=3pt,fill=vertexcolor_unnamed__1_1, draw=vertexbordercolor_unnamed__1,]
  \tikzstyle{vertexstyle_unnamed__1_11} = [text=black, inner sep=2pt, rectangle, rounded corners=3pt,fill=vertexcolor_unnamed__1_1, draw=vertexbordercolor_unnamed__1,]
  \tikzstyle{vertexstyle_unnamed__1_12} = [text=black, inner sep=2pt, rectangle, rounded corners=3pt,fill=vertexcolor_unnamed__1_1, draw=vertexbordercolor_unnamed__1,]
  \tikzstyle{vertexstyle_unnamed__1_13} = [text=black, inner sep=2pt, rectangle, rounded corners=3pt,fill=vertexcolor_unnamed__1_1, draw=vertexbordercolor_unnamed__1,]
  \tikzstyle{vertexstyle_unnamed__1_14} = [text=black, inner sep=2pt, rectangle, rounded corners=3pt,fill=vertexcolor_unnamed__1_1, draw=vertexbordercolor_unnamed__1,]
  \tikzstyle{vertexstyle_unnamed__1_15} = [text=black, inner sep=2pt, rectangle, rounded corners=3pt,fill=vertexcolor_unnamed__1_1, draw=vertexbordercolor_unnamed__1,]
  \tikzstyle{vertexstyle_unnamed__1_16} = [text=black, inner sep=2pt, rectangle, rounded corners=3pt,fill=vertexcolor_unnamed__1_1, draw=vertexbordercolor_unnamed__1,]
  \tikzstyle{vertexstyle_unnamed__1_17} = [text=black, inner sep=2pt, rectangle, rounded corners=3pt,fill=vertexcolor_unnamed__1_1, draw=vertexbordercolor_unnamed__1,]
  \tikzstyle{vertexstyle_unnamed__1_18} = [text=black, inner sep=2pt, rectangle, rounded corners=3pt,fill=vertexcolor_unnamed__1_1, draw=vertexbordercolor_unnamed__1,]
  \tikzstyle{vertexstyle_unnamed__1_19} = [text=black, inner sep=2pt, rectangle, rounded corners=3pt,fill=vertexcolor_unnamed__1_1, draw=vertexbordercolor_unnamed__1,]
  \tikzstyle{vertexstyle_unnamed__1_20} = [text=black, inner sep=2pt, rectangle, rounded corners=3pt,fill=vertexcolor_unnamed__1_1, draw=vertexbordercolor_unnamed__1,]
  \tikzstyle{vertexstyle_unnamed__1_21} = [text=black, inner sep=2pt, rectangle, rounded corners=3pt,fill=vertexcolor_unnamed__1_1, draw=vertexbordercolor_unnamed__1,]
  \tikzstyle{vertexstyle_unnamed__1_22} = [text=black, inner sep=2pt, rectangle, rounded corners=3pt,fill=vertexcolor_unnamed__1_1, draw=vertexbordercolor_unnamed__1,]
  \tikzstyle{vertexstyle_unnamed__1_23} = [text=black, inner sep=2pt, rectangle, rounded corners=3pt,fill=vertexcolor_unnamed__1_1, draw=vertexbordercolor_unnamed__1,]
  \tikzstyle{vertexstyle_unnamed__1_24} = [text=black, inner sep=2pt, rectangle, rounded corners=3pt,fill=vertexcolor_unnamed__1_1, draw=vertexbordercolor_unnamed__1,]
  \tikzstyle{vertexstyle_unnamed__1_25} = [text=black, inner sep=2pt, rectangle, rounded corners=3pt,fill=vertexcolor_unnamed__1_1, draw=vertexbordercolor_unnamed__1,]
  \tikzstyle{vertexstyle_unnamed__1_26} = [text=black, inner sep=2pt, rectangle, rounded corners=3pt,fill=vertexcolor_unnamed__1_1, draw=vertexbordercolor_unnamed__1,]
  \tikzstyle{vertexstyle_unnamed__1_27} = [text=black, inner sep=2pt, rectangle, rounded corners=3pt,fill=vertexcolor_unnamed__1_1, draw=vertexbordercolor_unnamed__1,]

  \definecolor{edgecolor_unnamed__1}{rgb}{ 0 0 0 }
  \tikzstyle{edgestyle_unnamed__1} = [thick,color=edgecolor_unnamed__1]


  \foreach \i/\k in {9/1,9/2,10/1,10/3,11/1,11/8,12/2,12/4,13/2,13/5,14/3,14/4,15/3,15/6,16/4,16/7,17/5,17/7,18/5,18/8,19/6,19/7,20/6,20/8,21/9,21/11,21/13,21/18,22/9,22/10,22/12,22/14,23/10,23/11,23/15,23/20,24/12,24/13,24/16,24/17,25/14,25/15,25/16,25/19,26/17,26/18,26/19,26/20,27/21,27/22,27/23,27/24,27/25,27/26} {
   \draw[edgestyle_unnamed__1] (v\i_unnamed__1) -- (v\k_unnamed__1);
  }

  \foreach \i/\label in 
  {
 1/1 3 5,
  2/3 4 5,
  3/1 5 6,
  4/4 5 6,
  5/2 3 4,
  6/1 2 6,
  7/2 4 6,
  8/1 2 3,
  9/3 5,
  10/1 5,
  11/1 3,
  12/4 5,
  13/3 4,
  14/5 6,
  15/1 6,
  16/4 6,
  17/2 4,
  18/2 3,
  19/2 6,
  20/1 2,
  21/3,
  22/5,
  23/1,
  24/4,
  25/6,
  26/2,
 27/ $\mathcal{P}_7$} {
    \node at (v\i_unnamed__1) [vertexstyle_unnamed__1_\i] {\label};
  }

\end{tikzpicture}
}
  \end{tikzpicture}
}
\end{center}
\caption{\label{HasseFig} Hasse diagram for the face lattice of the Newton polytope $\mathcal{P}_7$ \re{P7}. Compatibility of all cones of 
the dual fan can be read off from it.}
\end{figure}
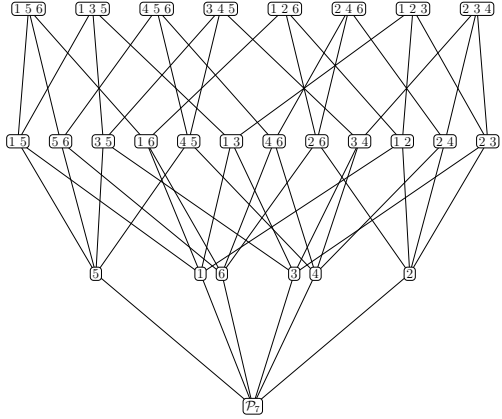

Let us proceed with a hands-on analysis. To construct the Newton polytope corresponding to the integrand \re{Region7Integrand}, we take the 
Minkowski sum of the polytopes of each of the integrand's polynomials, 
\begin{align}
\label{P7}
\mathcal{P}_7
\equiv
\mbox{Newt}[\mathcal{I}_7]
&
= \mbox{Newt}\, [x_1+1]  \oplus \mbox{Newt}\, [x_6 + x_7]  \oplus \mbox{Newt}\, [x_1 x_6 + x_1 x_7 + x_7 x_6 + x_1 x_7 x_6 ]
\nonumber\\
&
=
\mbox{ConvHull} \{
(2, 2, 1), (2, 2, 0), (2, 1, 2), (2, 1, 1), (2, 0, 2), (1, 2, 1), 
\nonumber\\
&\qquad\qquad\qquad
(1, 2, 0), (1, 1, 2), (1, 1, 1), (1, 0, 2), (0, 2, 1), (0, 1, 2) 
\}
\, .
\end{align}
The result is shown in Fig.\ \ref{NewtonPolytope}, where the $(t_1, t_2, t_3)$-axes label the exponents of $x_1$, $x_6$ and $x_7$, 
respectively. The polytope consists of 6 facets. The normals to these,---the rays $\bit{r}$ of its dual fan,---are shown with arrows 
(ignore the color-coding for now). They read explicitly
\begin{align} 
\label{RaysFanReg7}
\bit{r} = \{
(1,0,0),(0,1,0),(0,1,1),(0,-1,-1),(0,0,1),(-1,-1,-1)
\}
\, .
\end{align}
The facets are dual to these rays, and the vertices of the polytope are dual to maximal cones formed by triplets of the corresponding transversal 
facets' normals. Compatibility of cones of all dimensions can be read off from the face lattice of the polytope and is depicted in the Hasse 
diagram\footnote{Though most of the calculations in this section can be done in {\tt Mathematica} \cite{Mathematica}, we employed {\tt Polymake} 
\cite{polymake} as a main tool for polytope analyses in all subsequent considerations.} in Fig.\ \ref{HasseFig}. Consequences of the compatibility 
conditions will be discussed shortly.

The tropical approximation of the integrand ${\rm Trop} (\bit{t}) \equiv {\rm Trop} [\mathcal{I}_7]$ is given by the piece-wise multivariate linear function
\begin{align}
\mbox{Trop} (\bit{t})
=
&
-
(1 - 3 \ep) \max (0,t_1) 
\\
&
+ 3 \ep \max (t_2,t_3) - 2 \ep \max (t_1+t_2,t_1+t_3,t_2+t_3,t_1+t_2+t_3)
\, . \nonumber
\end{align}
Evaluating its value on the rays \re{RaysFanReg7}, we find that all but the first one are divergent,
\begin{align}
\mbox{Trop} (\bit{r}) = \{ - 1 + \ep, \ep, -\ep, -\ep, \ep, \ep \}
\, . 
\end{align}
So the divergent subfan 
\begin{align}
\label{DivFan7}
\Sigma_{\rm div} =
\{ 
&
{\rm Span}_+ (\bit{r}_2), {\rm Span}_+ (\bit{r}_3), {\rm Span}_+ (\bit{r}_4), {\rm Span}_+ (\bit{r}_5), {\rm Span}_+ (\bit{r}_6),
\\
&
{\rm Span}_+ (\bit{r}_2, \bit{r}_3), {\rm Span}_+ (\bit{r}_2, \bit{r}_4), {\rm Span}_+ (\bit{r}_2, \bit{r}_6) 
\nonumber\\
&
{\rm Span}_+ (\bit{r}_3, \bit{r}_4), {\rm Span}_+ (\bit{r}_3, \bit{r}_5),
\nonumber\\
&
{\rm Span}_+ (\bit{r}_4, \bit{r}_5), {\rm Span}_+ (\bit{r}_4, \bit{r}_6), 
\nonumber\\
&
{\rm Span}_+ (\bit{r}_5, \bit{r}_6), 
\nonumber\\
&
{\rm Span}_+ (\bit{r}_2, \bit{r}_3, \bit{r}_4),
{\rm Span}_+ (\bit{r}_2, \bit{r}_4, \bit{r}_6),
{\rm Span}_+ (\bit{r}_3, \bit{r}_4, \bit{r}_5),
{\rm Span}_+ (\bit{r}_4, \bit{r}_5, \bit{r}_6)
\} \nonumber
\end{align}
covers a large portion of the face lattice of $\mathcal{P}_7$ as can be seen from the corresponding Hasse graph in Fig.\ \ref{HasseFig}.

Our earlier estimates are pretty close to the real numbers stemming from the exercise in linear programming. There are five divergent rays 
in three dimensions. They are obviously linearly dependent: the divergent subfan \re{DivFan7} of the polytope $\mathcal{P}_7$ for region 7 is 
not simplicial, and the geometric property is not satisfied. To alleviate the problem with this overcomplete set of divergent rays, we need to 
get rid of `redundant' ones. There are, of course, a number of routes to achieve this. What drives the optimal choice is how many terms 
one generates after each step in the elimination process. This is heavily integrand-dependent.

\subsection{Nilsson-Passare IBP}

The main tool in the elimination of divergent rays is the Nilsson-Passare analytic continuation \cite{Nilsson,Berkesch} via integration by parts 
(IBP) formulated in the space of Schwinger parameters. Let us describe this procedure explicitly.

For the case at hand, we need to get rid of at least two divergent rays in the divergent subfan $\Sigma_{\rm div}$ \re{DivFan7} such that the rest
are linearly independent. A glance at Eqs.\ \re{RaysFanReg7} and \re{Region7Integrand} suggests that we should choose to integrate by parts 
first, along the divergent rays $\bit{r}_3 = (0,1,1)$ and $\bit{r}_4 = (0,-1,-1)$. One of the reasons for this choice is, of course, linear algebraic, 
and the other is that both of them produce just a single term after corresponding Nilsson-Passare procedures. For instance, to integrate along 
$\bit{r}_3$, we rescale the integration variables of $\mathcal{I}_7$ as $\bit{x} = (x_1, x_6, x_7) \to \bit{x}_\lambda \equiv \lambda^{-\bit{\scriptstyle r}_3} \bit{x} 
= (x_1, x_6/\lambda, x_7/\lambda)$ and, benefiting from the rescaling invariance of the logarithmic measure \re{Measure}, set
\begin{align}
\left. \frac{d \mathcal{I}_7 (\bit{x}_\lambda)}{d\lambda} \right|_{\lambda = 1} 
= 
0
\, .
\end{align}
In this manner, we relate the original integrand to a new one with a better behavior along the $\bit{r}_3$-ray,
\begin{align}
\mathcal{I}_7 (\bit{x}) = 2 \mathcal{I}_7^\prime (\bit{x})
\, , 
\end{align}
where
\begin{align}
\mathcal{I}_7^\prime (\bit{x}) 
= 
x_1 (1 + x_1)^{3 \ep} (x_6+x_7)^{1 + 3 \ep} (x_6 x_7 + x_1 x_6 + x_1 x_7 + x_1 x_6 x_7)^{- 1 - 2 \ep}/(1 + x_1/u)
\, .
\end{align}
With its tropical approximation being
\begin{align}
\mbox{Trop}^{\prime} (\bit{t})
&
=
t_1 - (1 - 3 \ep) \max (0,t_1) + (1 + 3 \ep) \max (t_2,t_3) 
\nonumber\\
&
- (1 + 2 \ep) \max (t_1+t_2,t_1+t_3,t_2+t_3,t_1+t_2+t_3)
\, ,
\end{align}
we find that the integral indeed becomes finite on the ray $\bit{r}_3$,
\begin{align}
\mbox{Trop}' (\bit{r}_3) = -1 - \ep
\, .
\end{align}
Since the factor-polynomial structure of the integrand did not change, we can now perform IBP along the ray $\bit{r}_4 = - \bit{r}_3$.
Of course, it also yields a single term,
\begin{align}
\mathcal{I}_7^\prime (\bit{x}) 
=
\frac{1 + 2 \ep}{\ep}
\mathcal{I}_7^{\prime\prime} (\bit{x}) 
\, ,
\end{align}
where
\begin{align}
\mathcal{I}_7^{\prime\prime} (\bit{x}) 
=
x_1 x_6 x_7 (1 + x_1)^{1 + 3 \ep} (x_6+x_7)^{1 + 3 \ep} (x_6 x_7 + x_1 x_6 + x_1 x_7 + x_1 x_6 x_7)^{- 2 - 2 \ep}/(1 + x_1/u)
\, .
\end{align}
With its Trop being 
\begin{align}
\mbox{Trop}^{\prime\prime} (\bit{t})
&
=
t_1 + t_2 + t_3 + 3 \ep \max (0,t_1) + (1 + 3 \ep)  \max (t_2,t_3) 
\nonumber\\
&
- 2(1 + \ep) \max (t_1+t_2,t_1+t_3,t_2+t_3,t_1+t_2+t_3)
\, ,
\end{align}
we find right away 
\begin{align}
\mbox{Trop}^{\prime\prime} (\bit{r}) = \{ - 1 + \ep,  \ep, - 1 - \ep, -1 - \ep, \ep, \ep \}
\, .
\end{align}
Since the values of Trops for this integrand are the same on all divergent rays, we will simply refer to these as $\mbox{Trop} \equiv 
\mbox{Trop}^{\prime\prime} (\bit{r}_{2,5,6}) = \ep$ in what follows. 

Having eliminated the two divergent rays $\bit{r}_{3,4}$, we are, in principle, done since the rest, i.e., $\bit{r}_{2,5,6}$, are linearly 
independent such that their fan is simplicial. The corresponding divergent subfan is formed by the cones
\begin{align}
\label{DivFanOpt1}
\Sigma^{\prime\prime}_{\rm div} 
= 
\{ 
\mbox{Span}_+ (\bit{r}_2), \mbox{Span}_+ (\bit{r}_5), \mbox{Span}_+ (\bit{r}_6), 
\mbox{Span}_+ (\bit{r}_2,\bit{r}_6), 
\mbox{Span}_+ (\bit{r}_5,\bit{r}_6)
\}
\, .
\end{align}
We will dub this as Option I below.

However, to demonstrate the variety of ways of handling the integral and make an informed decision on how to proceed in more
complex situations, we can do one more IBP such that the divergent subfan simplifies even further. Namely, we can integrate along 
either of the three remaining rays $\bit{r}_2$, $\bit{r}_5$, or $\bit{r}_6$. If we do IBP along $\bit{r}_6$, the remaining divergent 
rays $\bit{r}_{2,5}$ would be incompatible with respect to the normal fan, see Fig.\ \ref{HasseFig}. However, if we perform the 
Nilsson-Passare procedure along $\bit{r}_2$ or $\bit{r}_5$, the remaining rays will be compatible, and this will complicate the 
structure of the divergent subfan. Below, we will elucidate (dis)advantages for each of the chosen paths.

Let's apply the Nilsson-Passare procedure, say, for the ray $\bit{r}_2$. It immediately 
yields
\begin{align}
\mathcal{I}^{\prime\prime}_7 (\bit{x}) 
=
\frac{1 + 3 \ep}{\ep} \mathcal{I}^{\prime\prime\prime}_{7;1} (\bit{x}) 
-
2 \frac{1 + \ep}{\ep} \mathcal{I}^{\prime\prime\prime}_{7;2} (\bit{x}) 
\, ,
\end{align}
with
\begin{align}
\mathcal{I}^{\prime\prime\prime}_{7;1} (\bit{x}) 
&
=
x_1 x_6 x_7^2 (1 + x_1)^{1 + 3 \ep} (x_6+x_7)^{3 \ep} (x_6 x_7 + x_1 x_6 + x_1 x_7 + x_1 x_6 x_7)^{- 2 - 2 \ep}/(1 + x_1/u)
\, , \\
\mathcal{I}^{\prime\prime\prime}_{7;2} (\bit{x}) 
&
=
x_1^2 x_6 x_7^2 (1 + x_1)^{1 + 3 \ep} (x_6+x_7)^{1 + 3 \ep} (x_6 x_7 + x_1 x_6 + x_1 x_7 + x_1 x_6 x_7)^{- 1 - 2 \ep}/(1 + x_1/u)
\, .
\end{align}
For these two integrands, the divergent subfan consists of the two rays $\bit{r}_5$, $\bit{r}_6$, and their positive span,
\begin{align}
\label{DivFanOpt2}
\Sigma^{\prime\prime\prime}_{\rm div} = \{ \mbox{Span}_+ (\bit{r}_5), \mbox{Span}_+ (\bit{r}_6), \mbox{Span}_+ (\bit{r}_5,\bit{r}_6)\}
\, .
\end{align}
On the rest, they are finite, as can be seen from their Trops,
\begin{align}
\mbox{Trop}_1^{\prime\prime\prime} (\bit{t})
&
=
t_1 + t_2 + 2 t_3 + 3 \ep \max (0,t_1) + 3 \ep  \max (t_2,t_3) 
\nonumber\\
&
- 2(1 + \ep) \max (t_1+t_2,t_1+t_3,t_2+t_3,t_1+t_2+t_3)
\, , \\
\mbox{Trop}_2^{\prime\prime\prime} (\bit{t})
&
=
2 t_1 + t_2 + 2 t_3 + 3 \ep \max (0,t_1) + (1 + 3 \ep) \max (t_2,t_3) 
\nonumber\\
&
- (3 + 2 \ep) \max (t_1+t_2,t_1+t_3,t_2+t_3,t_1+t_2+t_3)
\, ,
\end{align}
with values
\begin{align}
\mbox{Trop}_1^{\prime\prime\prime} (\bit{r}) = \{ - 1 + \ep, -1 + \ep, - 1 - \ep, -1 - \ep, \ep, \ep \}
\, , \\
\mbox{Trop}_2^{\prime\prime\prime} (\bit{r}) = \{ - 1 + \ep, -1 + \ep, - 2 - \ep, -1 - \ep, \ep, \ep \}
\, .
\end{align}
Again, since both of the integrands possess the same values on both divergent rays, we will denote them as $\mbox{Trop} = 
\mbox{Trop}_1^{\prime\prime\prime} (\bit{r}_{5,6}) = \mbox{Trop}_2^{\prime\prime\prime} (\bit{r}_{5,6})= \ep$. The original integral 
is thus reduced to
\begin{align}
\mathcal{I}_7  (\bit{x}) 
= 
2 \frac{(1 + 2 \ep) (1 + 3 \ep)}{\ep^2} \mathcal{I}^{\prime\prime\prime}_{7;1}  (\bit{x}) 
-
4 \frac{(1 + \ep) (1 + 2 \ep)}{\ep^2} \mathcal{I}^{\prime\prime\prime}_{7;2}  (\bit{x}) 
\, ,
\end{align}
and we will refer to this as Option II.

Finally, if we choose instead to integrate along $\bit{r}_6$ as last IBP, we find three terms
\begin{align}
\mathcal{I}_7  (\bit{x}) 
= 
- 2 \frac{(1 + 2 \ep)}{\ep^2 u} \mathcal{I}^{\prime\prime\prime\prime}_{7;1}  (\bit{x}) 
+
2 \frac{(1 + 2 \ep) (1 + 3 \ep)}{\ep^2} \mathcal{I}^{\prime\prime\prime\prime}_{7;2}  (\bit{x}) 
-
4 \frac{(1 + \ep) (1 + 2 \ep)}{\ep^2} \mathcal{I}^{\prime\prime\prime\prime}_{7;3}  (\bit{x}) 
\, ,
\end{align}
with
\begin{align}
\mathcal{I}^{\prime\prime\prime\prime}_{7;1} (\bit{x}) 
&
=
x_1^2 x_6 x_7 (1 + x_1)^{1 + 3 \ep} (x_6+x_7)^{1 + 3 \ep} (x_6 x_7 + x_1 x_6 + x_1 x_7 + x_1 x_6 x_7)^{- 2 - 2 \ep}/(1 + x_1/u)^2
\, , \\
\mathcal{I}^{\prime\prime\prime\prime}_{7;2} (\bit{x}) 
&
=
x_1^2 x_6 x_7 (1 + x_1)^{3 \ep} (x_6+x_7)^{1 + 3 \ep} (x_6 x_7 + x_1 x_6 + x_1 x_7 + x_1 x_6 x_7)^{- 2 - 2 \ep}/(1 + x_1/u)
\, , \\
\mathcal{I}^{\prime\prime\prime\prime}_{7;3} (\bit{x}) 
&
=
x_1^2 x_6^2 x_7^2 (1 + x_1)^{1 + 3 \ep} (x_6+x_7)^{1 + 3 \ep} (x_6 x_7 + x_1 x_6 + x_1 x_7 + x_1 x_6 x_7)^{- 3 - 2 \ep}/(1 + x_1/u)
\, .
\end{align}
The divergent subfan for this situation is the simplest
\begin{align}
\label{DivFanOpt3}
\Sigma^{\prime\prime\prime\prime}_{\rm div} = \{ \mbox{Span}_+ (\bit{r}_2), \mbox{Span}_+ (\bit{r}_5) \}
\, ,
\end{align}
since the rays $\bit{r}_2$ and $\bit{r}_5$ are not compatible, see Fig.\ \ref{HasseFig}. This is Option III.

\subsection{Option I}

\begin{figure}[t]
\begin{center}
\mbox{
\begin{picture}(0,220)(120,0)
\put(0,0){\insertfig{8}{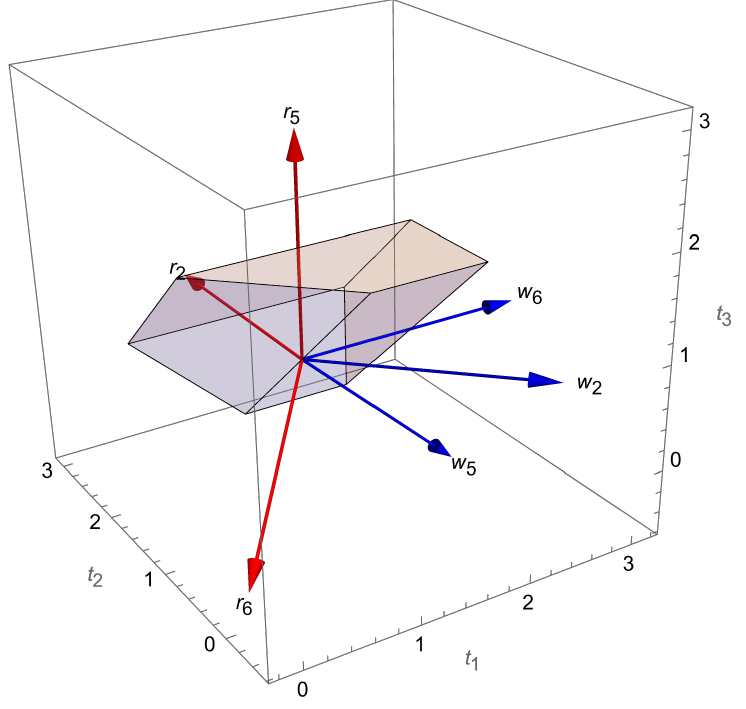}}
\end{picture}
}
\end{center}
\caption{\label{RaysandWsFig} Divergent rays $\bit{r}$ (red) and `conjugate' vectors $\bit{w}$ (blue) for $\Sigma^{\prime\prime}_{\rm div}$, Eq.\ 
\re{DivFanOpt1}, of the integrand $\mathcal{I}^{\prime\prime}_7$.}
\end{figure}

We continue our consideration by working out Option I, first. Its divergent subfan contains three rays, with one, i.e., $\bit{r}_6$, 
compatible both with $\bit{r}_2$ and $\bit{r}_5$, which makes the inclusion/exclusion a bit elaborate. We choose the conjugate vectors 
$\bit{w}_i$ according to the geometrical condition \re{GeometricProperty} to be
\begin{align}
\bit{w}_2 = (1,-1,0)
\, , \qquad
\bit{w}_5 = (1,0,-1)
\, , \qquad
\bit{w}_6 = (1,0,0)
\, .
\end{align}
In this manner, as can be seen from Fig.\ \ref{RaysandWsFig}, these do not {\sl refine} the divergent subfan. The $v$-variables \re{vVariables} 
corresponding to them are
\begin{align}
v_2 = \frac{1}{1 + x_1/x_6}
\, , \qquad
v_5 = \frac{1}{1 + x_1/x_7}
\, , \qquad
v_6 = \frac{1}{1 + x_1}
\, .
\end{align}
Then, the inclusion/exclusion allows us to write the integrand $\mathcal{I}^{\prime\prime}$ as a sum of six terms
\begin{align}
\mathcal{I}^{\prime\prime}_7
=
\sum_{i = 1}^6 \mathcal{I}^{\prime\prime (i)}_7
\, ,
\end{align}
where
\begin{align}
\mathcal{I}^{\prime\prime (1)}_7
&
= \mathcal{I}^{\prime\prime} _7
- 
v_2^{1 - {\rm Trop}} 
\left. \mathcal{I}^{\prime\prime}_7 \right|_{\bit{\scriptstyle r}_2} 
- 
v_5^{1 - {\rm Trop}} 
\left. \mathcal{I}^{\prime\prime}_7 \right|_{\bit{\scriptstyle r}_5}
- 
v_6^{1 - {\rm Trop}} 
\left. \mathcal{I}^{\prime\prime}_7 \right|_{\bit{\scriptstyle r}_6}
\\
&
+
v_2^{1 - {\rm Trop}} v_6^{1 - {\rm Trop}} 
\left. \mathcal{I}^{\prime\prime}_7 \right|_{{\rm Span}_+ (\bit{\scriptstyle r}_2, \bit{\scriptstyle r}_6)} 
+
v_5^{1 - {\rm Trop}} v_6^{1 - {\rm Trop}} 
\left. \mathcal{I}^{\prime\prime}_7 \right|_{{\rm Span}_+ (\bit{\scriptstyle r}_5, \bit{\scriptstyle r}_6)} 
\, , \nonumber\\
\mathcal{I}^{\prime\prime (2)}_7
&
= 
v_2^{1 - {\rm Trop}} 
\left[
\left. \mathcal{I}^{\prime\prime}_7 \right|_{\bit{\scriptstyle r}_2} 
- 
v_6^{1 - {\rm Trop}} 
\left. \mathcal{I}^{\prime\prime}_7 \right|_{{\rm Span}_+ (\bit{\scriptstyle r}_2, \bit{\scriptstyle r}_6)} 
\right]
\, , \\
\mathcal{I}^{\prime\prime (3)}_7 
&
= 
v_5^{1 - {\rm Trop}} 
\left[
\left. \mathcal{I}^{\prime\prime}_7 \right|_{\bit{\scriptstyle r}_5} 
- 
v_6^{1 - {\rm Trop}} 
\left. \mathcal{I}^{\prime\prime}_7 \right|_{{\rm Span}_+ (\bit{\scriptstyle r}_5, \bit{\scriptstyle r}_6)} 
\right]
\, , \\
\mathcal{I}^{\prime\prime (4)}_7
&
= 
v_6^{1 - {\rm Trop}} 
\left[
\left. \mathcal{I}^{\prime\prime}_7 \right|_{\bit{\scriptstyle r}_6} 
- 
v_2^{1 - {\rm Trop}} 
\left. \mathcal{I}^{\prime\prime}_7 \right|_{{\rm Span}_+ (\bit{\scriptstyle r}_2, \bit{\scriptstyle r}_6)} 
- 
v_5^{1 - {\rm Trop}} 
\left. \mathcal{I}^{\prime\prime}_7 \right|_{{\rm Span}_+ (\bit{\scriptstyle r}_5, \bit{\scriptstyle r}_6)} 
\right]
\, , \\
\mathcal{I}^{\prime\prime (5)}_7
&
= 
v_2^{1 - {\rm Trop}} 
v_6^{1 - {\rm Trop}} 
\left. \mathcal{I}^{\prime\prime}_7 \right|_{{\rm Span}_+ (\bit{\scriptstyle r}_2, \bit{\scriptstyle r}_6)} 
\, , \\
\mathcal{I}^{\prime\prime (6)}_7 
&
= 
v_5^{1 - {\rm Trop}} 
v_6^{1 - {\rm Trop}} 
\left. \mathcal{I}^{\prime\prime}_7 \right|_{{\rm Span}_+ (\bit{\scriptstyle r}_5, \bit{\scriptstyle r}_6)} 
\, .
\end{align}
Here, the initial forms are
\begin{align}
&
\left. \mathcal{I}_7^{\prime\prime} (\bit{x}; u) \right|_{\bit{\scriptstyle r}_2} 
=
x_1 x_6^{\ep} x_7 (1 + x_1)^{3 \ep+1} (x_1 + x_7 + x_1 x_7)^{-2 \ep-2}/(1 + x_1/u)
\, , \\
&
\left. \mathcal{I}_7^{\prime\prime} (\bit{x}; u) \right|_{\bit{\scriptstyle r}_5} 
=
x_1 x_6 x_7^{\ep} (1 + x_1)^{3 \ep+1} (x_1 + x_6 + x_1 x_6)^{-2 \ep-2}/(1 + x_1/u)
\, , \\
&
\left. \mathcal{I}_7^{\prime\prime} (\bit{x}; u) \right|_{\bit{\scriptstyle r}_6} 
=
x_1 x_6 x_7 (x_6 + x_7)^{3 \ep+1} (x_1 x_6 + x_1 x_7 + x_6 x_7)^{-2 \ep-2}
\, , \\
&
\left. \mathcal{I}_7^{\prime\prime} (\bit{x}; u) \right|_{{\rm Span}_+ (\bit{\scriptstyle r}_2,\bit{\scriptstyle r}_6)} 
=
x_1 x_6^\ep x_7 (x_1 + x_7)^{-2 \ep - 2}
\, , \\
&
\left. \mathcal{I}_7^{\prime\prime} (\bit{x}; u) \right|_{{\rm Span}_+ (\bit{\scriptstyle r}_5,\bit{\scriptstyle r}_6)} 
=
x_1 x_6 x_7^\ep (x_1 + x_6)^{-2 \ep - 2}
\, .
\end{align}

Let us now count how many terms we need to calculate, taking into account the Laurent expansion. Due to an overall double
pole in $\ep$ accompanying $\mathcal{I}_7^{\prime\prime}$, we need to expand $\mathcal{I}_7^{(1)}$ to $o (\ep^2)$, while 
$\mathcal{I}_7^{(2,3,4)}$ to $o (\ep^3)$ after extracting from them an overall pole in $\ep$. Finally, $\mathcal{I}^{(5,6)}$ can
be calculated exactly. Were we not able to determine the latter two exactly, they would contribute 5 terms each. Adding these 
up, there are 17 (or 27, with the last stipulation) terms total to be fed to {\tt HyperInt}.

\subsection{Option II}

For this calculation, the divergent fan \re{DivFanOpt2} consists of two {\sl compatible} rays. We can choose the `conjugate' 
vectors as
\begin{align}
\bit{w}_5 = (0,1,-1)
\, , \qquad
\bit{w}_6 = (0,1,0)
\, ,
\end{align}
with the corresponding $v$-variables being
\begin{align}
v_5 = \frac{1}{1 + x_6/x_7}
\, , \qquad
v_6 = \frac{1}{1 + x_6}
\, .
\end{align}
Then, the inclusion/exclusion recasts the integrands $\mathcal{I}^{\prime\prime\prime}_{1,2}$, cumulatively called $\mathcal{I}^{\prime\prime\prime}$ 
below, as a sum of four terms
\begin{align}
\mathcal{I}^{\prime\prime\prime}
=
\sum_{i = 1}^4
\mathcal{I}^{\prime\prime\prime(i)}
\, ,
\end{align}
where
\begin{align}
\mathcal{I}^{\prime\prime\prime(1)} 
&
= \mathcal{I}^{\prime\prime\prime} 
- 
v_5^{1 - {\rm Trop}} 
\left. \mathcal{I}^{\prime\prime\prime} \right|_{\bit{\scriptstyle r}_5} 
- 
v_6^{1 - {\rm Trop}} 
\left. \mathcal{I}^{\prime\prime\prime} \right|_{\bit{\scriptstyle r}_6}
+
v_5^{1 - {\rm Trop}} v_6^{1 - {\rm Trop}} 
\left. \mathcal{I}^{\prime\prime\prime} \right|_{{\rm Span}_+ (\bit{\scriptstyle r}_5, \bit{\scriptstyle r}_6)} 
\, , \\
\mathcal{I}^{\prime\prime\prime(2)} 
&
= 
v_5^{1 - {\rm Trop}} 
\left[
\left. \mathcal{I}^{\prime\prime\prime} \right|_{\bit{\scriptstyle r}_5} 
- 
v_6^{1 - {\rm Trop}} 
\left. \mathcal{I}^{\prime\prime\prime} \right|_{{\rm Span}_+ (\bit{\scriptstyle r}_5, \bit{\scriptstyle r}_6)} 
\right]
\, , \\
\mathcal{I}^{\prime\prime\prime(3)} 
&
= 
v_6^{1 - {\rm Trop}} 
\left[
\left. \mathcal{I}^{\prime\prime\prime} \right|_{\bit{\scriptstyle r}_6} 
- 
v_5^{1 - {\rm Trop}} 
\left. \mathcal{I}^{\prime\prime\prime} \right|_{{\rm Span}_+ (\bit{\scriptstyle r}_5, \bit{\scriptstyle r}_6)} 
\right]
\, , \\
\mathcal{I}^{\prime\prime\prime(4)} 
&
= 
v_5^{1 - {\rm Trop}} 
v_6^{1 - {\rm Trop}} 
\left. \mathcal{I}^{\prime\prime\prime} \right|_{{\rm Span}_+ (\bit{\scriptstyle r}_5, \bit{\scriptstyle r}_6)} 
\end{align}
are built in terms of the following initial forms for the first
\begin{align}
&
\left. \mathcal{I}^{\prime\prime\prime}_1 (\bit{x}) \right|_{\bit{\scriptstyle r}_5} 
=
x_1 x_6 x_7^{\ep} (1 + x_1)^{3 \ep+1} (x_1 + x_6 + x_1 x_6)^{-2 \ep-2}/(1 + x_1/u)
\, , \\
&
\left. \mathcal{I}^{\prime\prime\prime}_1 (\bit{x}) \right|_{\bit{\scriptstyle r}_6} 
=
x_1 x_6 x_7^2 (x_6 + x_7)^{3 \ep+1} (x_1 x_6 + x_1 x_7 + x_6 x_7)^{- 2 \ep - 2}
\, , \\
&
\left. \mathcal{I}^{\prime\prime\prime}_1 (\bit{x}) \right|_{{\rm Span}_+ (\bit{\scriptstyle r}_5, \bit{\scriptstyle r}_6)} 
=
x_1 x_6 x_7^\ep (x_1 + x_6)^{- 2 \ep - 2}
\, , 
\end{align}
and second 
\begin{align}
&
\left. \mathcal{I}^{\prime\prime\prime}_2 (\bit{x}) \right|_{\bit{\scriptstyle r}_5} 
=
x_1^2 x_6 x_7^{\ep} (x_1+1)^{3 \ep+1} (x_1 + x_6 + x_1 x_6)^{-2 \ep-3} /(1 + x_1/u)
\, , \\
&
\left. \mathcal{I}^{\prime\prime\prime}_2 (\bit{x}) \right|_{\bit{\scriptstyle r}_6} 
=
x_1^2 x_6 x_7^2 (x_6+x_7)^{3 \ep+1} (x_1 x_6 + x_1x_7 + x_6 x_7)^{-2 \ep-3}
\, , \\
&
\left. \mathcal{I}^{\prime\prime\prime}_1 (\bit{x}) \right|_{{\rm Span}_+ (\bit{\scriptstyle r}_5, \bit{\scriptstyle r}_6)} 
=
x_1^2 x_6 x_7^{\ep} (x_1+x_6)^{-2 \ep-3}
\, .
\end{align}
contributions, respectively.

Now, we are back to the counting of how many terms, including the Laurent expansion, we need in this case. Since we had one extra IBP along a
divergent ray, $\mathcal{I}^{\prime\prime\prime}$'s are accompanied by the third-order pole in $\ep$. So, $\mathcal{I}^{\prime\prime\prime(1)}$ 
needs to be expanded to $o (\ep^3)$,  $\mathcal{I}^{\prime\prime\prime(2,3)}$ to $o (\ep^4)$ after the separation of the overall divergence, and 
$\mathcal{I}^{\prime\prime\prime(4)}$ is known exactly. If the latter were not possible, it would produce 5 terms instead. We thus have to calculate 
$2\times (3 + 2 \times 4 + 1) = 24$, respectively, 28 terms.

\subsection{Option III}

Last but not least, let us address Option III. Its divergent subfan \re{DivFanOpt3} consists of just two {\sl incompatible} rays. Thus, the 
required `subtractions' are extremely simple, and the inclusion/exclusion merely contains three terms
\begin{align}
\mathcal{I}^{\prime\prime\prime\prime} 
= 
\mathcal{I}^{\prime\prime\prime\prime(1)} + \mathcal{I}^{\prime\prime\prime\prime(2)} + \mathcal{I}^{\prime\prime\prime\prime(3)} 
\end{align}
for each of $\mathcal{I}^{\prime\prime\prime\prime} = \mathcal{I}^{\prime\prime\prime\prime}_{1,2,3}$, where
\begin{align}
&
\mathcal{I}^{\prime\prime\prime\prime(1)} 
= \mathcal{I}^{\prime\prime\prime\prime}
- 
v_2^{1 - {\rm Trop}} 
\left. \mathcal{I}^{\prime\prime\prime\prime} \right|_{\bit{\scriptstyle r}_2} 
- 
v_5^{1 - {\rm Trop}} 
\left. \mathcal{I}^{\prime\prime\prime\prime} \right|_{\bit{\scriptstyle r}_5}
\, , \\
&
\mathcal{I}^{\prime\prime\prime\prime(2)} = v_2^{1 - {\rm Trop}}  \left. \mathcal{I}^{\prime\prime\prime\prime} \right|_{\bit{\scriptstyle r}_2} 
\, , \\
&
\mathcal{I}^{\prime\prime\prime\prime(3)} = v_5^{1 - {\rm Trop}}  \left. \mathcal{I}^{\prime\prime\prime\prime} \right|_{\bit{\scriptstyle r}_5} 
\, .
\end{align}
Choosing the `conjugate' vectors in the form 
\begin{align}
\bit{w}_6 = (0,-1,0)
\, , \qquad
\bit{w}_7 = (0,0,-1)
\end{align} 
the $v$-variables read
\begin{align}
v_6 = \frac{1}{1 + 1/x_6}
\, , \qquad
v_7 = \frac{1}{1 + 1/x_7}
\, ,
\end{align}
and the initial forms are
\begin{align}
&
\left. \mathcal{I}^{\prime\prime\prime\prime}_1 (\bit{x}) \right|_{\bit{\scriptstyle r}_2} 
=
x_1^2 x_6^\ep x_7 (1 + x_1)^{3 \ep+1} (x_1 + x_7 + x_1 x_7)^{-2 \ep-2}/(1 + x_1/u)^2
\, , \\
&
\left. \mathcal{I}^{\prime\prime\prime\prime}_1 (\bit{x}) \right|_{\bit{\scriptstyle r}_5} 
=
x_1^2 x_6 x_7^\ep (1 + x_1)^{3 \ep+1} (x_1 + x_6 + x_1 x_6)^{-2 \ep-2}/(1 + x_1/u)^2
\, , \\
&
\left. \mathcal{I}^{\prime\prime\prime\prime}_2 (\bit{x}) \right|_{\bit{\scriptstyle r}_2} 
=
x_1^2 x_6^\ep x_7 (1 + x_1)^{3 \ep} (x_1 + x_7 + x_1 x_7)^{-2 \ep-2}/(1 + x_1/u)
\, , \\
&
\left. \mathcal{I}^{\prime\prime\prime\prime}_2 (\bit{x}) \right|_{\bit{\scriptstyle r}_5} 
=
x_1^2 x_6 x_7^\ep (1 + x_1)^{3 \ep} (x_1 + x_7 + x_1 x_7)^{-2 \ep-2}/(1 + x_1/u)
\, , \\
&
\left. \mathcal{I}^{\prime\prime\prime\prime}_3 (\bit{x}) \right|_{\bit{\scriptstyle r}_2} 
=
x_1^2 x_6^\ep x_7^2 (1 + x_1)^{3 \ep + 1} (x_1 + x_7 + x_1 x_7)^{-2 \ep-3}/(1 + x_1/u)
\, , \\
&
\left. \mathcal{I}^{\prime\prime\prime\prime}_3 (\bit{x}) \right|_{\bit{\scriptstyle r}_5} 
=
x_1^2 x_6^2 x_7^\ep (1 + x_1)^{3 \ep + 1} (x_1 + x_6 + x_1 x_6)^{-2 \ep-3}/(1 + x_1/u)
\, .
\end{align}

Adding up the multiplicities of all ingredients involved, we conclude that we need $3(3+4+4) = 33$ terms to calculate.
There are more than in the previous two options. However, this is not the most unpleasant feature. In our case of many kinematical 
variables, the divergent subfan for this particular option results in spurious kinematical poles in individual contributions. We find 
them `crippled' with the $(1-u)$-denominator, which disappears only in the total sum. While this can be looked at as a feature 
rather than a bug, i.e., their cancellation ensures the correctness of the final result, in more complicated circumstances, this is 
an unwelcome characteristic of this choice. We observed in other contributions from regions that we will face a dilemma 
in choosing either dealing with spurious poles in smaller expressions or rather without them, but with factor-of-ten larger 
integration outputs. The take-home lesson of our consideration in this section is that the simpler the divergent fan is, the 
more terms one has to calculate! So, it is a trade-off. Of course, no matter the chosen route, the final integration step with 
{\tt HyperInt} gives the same result. However, there are subtleties. We address them next.

\section{Breaking of linear reducibility}
\label{LinearRedSection}

Up to now, we have not been very verbose about the integration step of the formalism. Although the {\tt Hyperint}'s integrator
is very powerful, its {\sl responses are limited, you must ask the right questions} \cite{irobotquote}. The main problem that
arises is the violation of the linear reducibility of integrands after a few integration steps have already been taken. If this 
happens after all but one remaining integration, {\tt HyperInt} can be instructed \cite{Panzer:2014gra} to work with algebraic 
roots by enabling the option 
\begin{equation}
\verb|_hyper_algebraic_roots:=true|
\end{equation}
and explicitly specifying the field to factorize quadratic polynomials which yields these radicals
\begin{equation}
\label{SplitFieldEq}
\verb|_hyper_splitting_field:={RootOf(c+b*_Z+a*_Z^2)}|
\, .
\end{equation}
However, if there is more than one variable resulting in quadratic polynomials, the only way out is to implement a proper variable 
transformation before the next step in the integration recursion can be taken. There is no general prescription for how to do it, at 
least for the moment. So, it has to be figured out on a case-by-case basis. 

Let us demonstrate this with an example of a region where this is encountered. We will take the most involved integral as far 
as the number of integrations is concerned, i.e., region 4 with seven Schwinger parameters before fixing the ${\rm GL} (1)$-redundancy. 
First, we want to reduce the number of integrations to be performed. A quick inspection of the integrand \re{Reg4Init} suggests that 
we can integrate over the sum $\eta = x_7 + x_8$ by first implementing the change of variable $x_7 = \eta \xi$, $x_8 = 
\eta (1 - \xi)$ with the Jacobian $\eta$. Then we choose the gauge $x_2 = 1$. Finally, to work with all integrations over the positive
half line, we further replace $\xi = x_2/(1+x_2)$ with the Jacobian $1/(1+x_2)^2$. These manipulations simplify this region
to a sum of three five-fold integrals
\begin{align}
\label{Reg4Init2}
{\rm Reg}_4 
= 
c_4 \int \frac{d^5 \bit{x}}{\bit{x}} 
\big[ (2 + \ep) s_{51} \mathcal{I}_{4,a} (\bit{x}) + (2 + \ep) s_{34} \mathcal{I}_{4,b} (\bit{x}) - 2 \ep \mathcal{I}_{4,c} (\bit{x}) \big]
\, ,
\end{align}
with 
\begin{align}
c_4 
= \frac{\e^{2 \gamma_{\rm E}  \ep}}{s_{45}} \Gamma (\ep) \Gamma (2 + \ep)
\end{align}
and the integrands
\begin{align}
&
\mathcal{I}_{4,a} (\bit{x})
=
x_3 \mathcal{I}_{4,b} (\bit{x})
=
x_1 x_2^{1-\ep}x_3^2 x_4 x_5  
(1 + x_2)^{2 + 3 \ep} (1 +x_1+x_3+x_4+x_5)^{2 \ep} 
\\
&\times
\left[
s_{12} (x_1 x_4+x_4+x_5)
\!+\!
s_{23} (1 + x_1) x_2 x_3
\!+\!
s_{34} x_1 x_5
\!+\!
s_{45} x_2 (x_1 x_4+x_4+x_5)
\!+\!
s_{51} x_1 x_5 x_3
\right]^{-3 - \ep}
\, , \nonumber\\
&
\mathcal{I}_{4,c} (\bit{x})
=
x_1 x_2^{1-\ep} x_3 x_4 x_5  
(1 + x_2)^{1 + 3 \ep} (1+x_1+x_3+x_4+x_5)^{-1 + 2 \ep} 
\\
&\times
\left[
s_{12} (x_1 x_4+x_4+x_5)
\!+\!
s_{23} (1 + x_1) x_2 x_3
\!+\!
s_{34} x_1 x_5
\!+\!
s_{45} x_2 (x_1 x_4+x_4+x_5)
\!+\!
s_{51} x_1 x_5 x_3
\right]^{-2 - \ep}
\, . \nonumber
\end{align}
The reader may wonder why we split $\mathcal{I}_{4,a} (\bit{x})$ and $\mathcal{I}_{4,b} (\bit{x})$ individually instead of considering them 
together. Notice that they differ from each other only by a power of $x_3$. The reason for this is that while overall monomials do not affect
individual Newton polytopes for each of these integrands, in the Minkowski sum of the two, one is shifted in the ambient space with respect to 
the other one by one lattice space in the $x_3$-direction. This adds extra facets for the total sum, actually two of them to be precise. If neither
were divergent, we could equally well consider the sum. However, one of them is, so this results in a more complicated divergent subfan. We 
chose to deal with simpler ones but doubled the number of terms to be computed. Again, it is a trade-off in effort. Currently, we do not have the
intention to be optimal.

We will address the simpler integrand $\mathcal{I}_{4,c}$ since it showcases integration pitfalls transparently. It needs to be evaluated merely 
to leading order in $\ep$, i.e., $o (\ep^0)$ since the pole in $c_4$ is cancelled by the accompanying power of $\ep$ in \re{Reg4Init2}.
The analysis of its Newton polytope with {\tt Polymake} shows that it possesses 13 facets, but just one of them is associated with a 
divergent ray in its dual fan, i.e., $\bit{r} = (0,1,0,0,0)$ with the value of
\begin{align}
{\rm Trop} (\bit{t})
&=
 t_1 + (1-\ep) t_2 +t_3 + t_4 + t_5
 \\
&
+
(1 + 3 \ep) \max (0,t_2)
-
(1-2 \ep) \max 
(0,
t_1,
t_3,
t_4,
t_5)
\nonumber\\
-
(2&+\ep) 
\max (
t_1 \!+\! t_3,
t_1 \!+\! t_2 \! +\! t_3,
t_4,
t_1\! +\! t_4,
t_2\! +\! t_4,
t_1\! +\! t_2\! +\! t_4,
t_5,
t_1\! +\! t_5,
t_2\! +\! t_5,
t_2\! +\! t_3\! +\! t_5
)
\, , \nonumber
\end{align}
being ${\rm Trop} (\bit{r}) = \ep$. The inclusion/exclusion formula contains just two terms
\begin{align}
\label{IntegrandI4}
\mathcal{I}_{4,c} (\bit{x})
=
\left[
\mathcal{I}_{4,c} - v_{\bit{\scriptstyle r}}^{1 - {\rm Trop} (\bit{\scriptstyle r})} \mathcal{I}_{4,c}|_{\bit{\scriptstyle r}}
\right]
+
v_{\bit{\scriptstyle r}}^{1 - {\rm Trop} (\bit{\scriptstyle r})} \mathcal{I}_{4,c}|_{\bit{\scriptstyle r}}
\, ,
\end{align}
where the subtraction involves the $v$-variable $v_{\bit{\scriptstyle r}} = x_2/(1 + x_2)$ and the initial form on the ray is
\begin{align}
\mathcal{I}_{4,c}|_{\bit{\scriptstyle r}}
&
=
x_1 x_2^{\ep} x_3 x_4 x_5
(1+x_1+x_3+x_4+x_5)^{-1 + 2 \ep}
\\ 
&\times
[s_{12} x_4+s_{23} x_1 x_3+s_{34} x_5+s_{45} x_1 x_4+s_{51} x_5 x_3]^{-2-\ep}
\, . \nonumber
\end{align}
The divergent integration in the `counterterm' can be performed using the barycentric variable transformation with the matrix 
$\bit{T} = (\bit{r}|\bit{e}_1|\bit{e}_3|\bit{e}_4|\bit{e}_5)$ formed by the divergent ray $\bit{r}$ and the five-dimensional unit vectors  
$\bit{e}_j$ along the $j$-th axis. The rest of the $\bit{y} = (y_2, y_3, y_4, y_5)$-integrals have to be fed to {\tt HyperInt}.

We turn to it next. No matter the order of integrations we choose, we end up with quadratic polynomials in the integrand after the
first two steps are executed. The property of linear reducibility is broken! However, {\tt HyperInt}'s error-message output specifies the 
first quadratic polynomial it encountered, so we can use this information to fix the problem. Of course, if there is more than one, 
the procedure will have to be repeated. For the `counterterm' in question, we chose the integration order $y_5 \to y_4 \to y_3 \to y_2$. 
After the $y_{5,4}$-integrations, the `non-factorizable' quadratic polynomial that arises in the integrand is 
$$
s_{23} y_2 y_3 - s_{34} (1 + y_2 + y_3) - s_{51} y_3 (1 + y_2 + y_3)
\, .
$$ 
It is almost obvious that one can linearize the $y_3$-integration with the variable change $y_3 \to (1+y_2) y_3$. This does the trick.
The final integration with respect to $y_2$ will, however, be quadratic as well. But, explicitly specifying the radicals in the option 
\re{SplitFieldEq} with 
\begin{align*}
a &= s_{45} s_{51} - s_{12} s_{51} - s_{23} s_{51}
\, , \\
b &= s_{12} s_{23} - s_{12} s_{51} - s_{23} s_{34} + s_{34} s_{45} + s_{45} s_{51}
\, , \\
c &= s_{34} s_{45}
\, ,
\end{align*}
{\tt HyperInt} performs the integration successfully.

The locally finite integrand in the first term in the square brackets of \re{IntegrandI4} can be done with the very same routine, except for
adding one more integration. This does not present a complication for {\tt HyperInt}.

\section{Adding up all contributions of regions}
\label{ResultsSection}

Strategies outlined in the previous sections were applied with minor variations to all other regions. We observed the appearance of other radicals
that produced other square-root letters in the pentabox's alphabet. We borrowed\footnote{We are thankful to Dima Chicherin for this.} the hard 
region, region 3, from the known results for pentabox Master Integrals on the conformal branch of the $\mathcal{N} = 4$ sYM
\cite{Gehrmann:2018yef,Papadopoulos:2015jft,Chicherin:2020oor}. Finally, we added up all regions with accompanying scalings in $m$ 
(\ref{PentaboxRegionsRep}). As expected, all poles in $\ep$ cancelled in the total expression since infrared singularities are regularized with 
the nonzero $m$. The pentabox enjoys the anticipated double logarithmic nature in $m$, such that it admits the following expansion at this 
perturbative order
\begin{align}
s_{12} s_{23} s_{45} G_{11111111-100}|_{m \to 0}
=
\sum_{\ell = 0}^4 a_\ell \log^{4-\ell} m^2
\, .
\end{align}
The leading three coefficients are simple enough to be presented here. They are
\begin{alignat}{2}
&
a_0 
&&
= 3
\, , \\
&
a_1
&&
=
- \frac{5}{2} \log (s_{12} s_{23} s_{34} s_{51}) - 2 \log (s_{45})
\, , \\
&
a_2
&&
=
\frac{1}{2} \log^2 \left(\frac{s_{34}}{s_{51}} \right)
+
\frac{1}{4} \log^2\left(\frac{s_{12}}{s_{23}}\right)
-
\frac{5}{4} \log^2(s_{12} s_{23})
+
\frac{3}{2} \log ^2(s_{34} s_{51})
\\
&
&&
+
\frac{3}{2} \log (s_{12}) \log(s_{34})
+
\frac{9}{2} \log (s_{12}) \log (s_{51})
+
\frac{9}{2} \log (s_{23}) \log (s_{34})
\nonumber\\
&
&&
+
\frac{3}{2} \log (s_{23}) \log (s_{51})
-
\frac{1}{2} \log (s_{45}) \log (s_{34} s_{51})
+
\frac{13}{2} \log (s_{45}) \log (s_{12} s_{23})
\nonumber\\
&
&&
-
4 \log (s_{34}) \log (s_{51})
-
3 \log ^2(s_{45})
+
\frac{5 \pi ^2}{6}
\, . \nonumber
\end{alignat}
The rest of the coefficients are not as simple, at least at this time. Therefore, they are relegated to the accompanying {\tt Mathematica}
notebook {\tt pentabox.nb}, where we also perform numerical tests of our results in terms of GPLs by means of the {\tt ginsh} interface
of {\tt PolyLogTools} \cite{Duhr:2019tlz} against sector decomposition with {\tt FIESTA}.

\section{Conclusions}

In this paper, we calculated a two-loop Feynman integral that eluded traditional techniques employed by practitioners in the field of 
multiloop computations of Feynman integrals. The formalism that allowed us to break through the barrier of limitations of previously
available methods is based on the tropical geometry of convex polytopes associated with parametric integrands. Since the integrals 
are infrared divergent, in order to be able to use powerful integrators available in the literature, it was imperative to deal with finite 
integrals by introducing subtractions {\sl under} the integral, which yield {\sl locally finite} integrands. Such a technique has been 
proposed in Ref.\ \cite{Salvatori:2024nva}. We relied on it as a blueprint for our consideration in this work. 

We successfully evaluated the near mass-shell limit of the pentabox in terms of GPLs. This was the last required ingredient for 
understanding the infrared properties of five-leg amplitudes at two-loop order on the Coulomb branch of $\mathcal{N} = 4$ sYM. 
This analysis is, however, postponed to a future publication.

Our consideration opens new avenues for further studies. It is likely that the same technique will allow us to go to even higher
multiplicities. Six-leg amplitudes are our ultimate goal. The main interest in them is to understand their finite part, since it will be the 
first occurrence of the so-called remainder functions of conformal cross-ratios formed by Mandelstam-like invariants. Are they the 
same as the ones in the conformal case \cite{Bern:2008ap,Drummond:2008aq,Dixon:2011nj}? If they are different, what can the 
difference be attributed to? In similar circumstances of near mass-shell form factors, an analogue of the remainder function shows 
up for a lower number of external legs. It was found there \cite{Belitsky:2024agy,Belitsky:2024dcf}, that it is indeed different from the 
conformal case \cite{Brandhuber:2012vm,Brandhuber:2014ica}, and the reason for this is a different form of splitting amplitudes which 
drive their near-collinear behavior \cite{Belitsky:2024rwv,Belitsky:2025pnw}. 

To achieve the above goal, an automation of the steps involved in the construction of convex polytopes, their tropical data, locally finite 
integrands, and, eventually, their integration is highly desirable. A {\tt Mathematica} package is being prepared to meet these needs.

\begin{acknowledgments}
We are grateful to Dima Chicherin for his generous help with the retrieval of the massless pentabox expressions.
We are deeply indebted to Giulio Salvatori for explanations of the formalism in Ref.\ \cite{Salvatori:2024nva}. The work of A.B. was supported by 
the U.S.\ National Science Foundation under grant No.\ PHY-2207138. The work of V.S. was conducted under the state assignment of Lomonosov 
Moscow State University and supported by the Moscow Center for Fundamental and Applied Mathematics of Lomonosov Moscow State University 
under Agreement No.\ 075-15-2025-345.
\end{acknowledgments}

\appendix

\section{Region integrals and their symmetries}
\label{RegionIntegrals}

In this appendix, we quote the Feynman/Schwinger integrals for all regions. They are enumerated according to Eq.\ \re{PentaboxRegionsRep} 
and have the following region vectors\footnote{Recall that we are using outward pointing normals as compared to the conventions adopted in 
{\tt asy}.}
\begin{align}
- \bit{r} 
= 
\{
&
\{1,0,1,2,2,1,0,1\}
,
\{2,1,0,1,2,1,0,1\}
,
\{0,0,0,0,0,0,0,0\}
,
\nonumber\\
&
\{1,1,1,1,1,1,0,0\}
,
\{1,0,1,2,1,0,1,2\}
,
\{2,1,0,1,1,2,1,0\}
,
\nonumber\\
&
\{0,0,1,1,1,0,0,1\}
,
\{2,1,0,1,2,1,1,2\}
,
\{1,0,1,2,1,1,2,2\}
,
\nonumber\\
&
\{1,1,0,0,1,0,0,1\}
,
\{1,1,2,2,2,1,0,1\}
,
\{1,0,1,2,2,1,1,2\}
,
\nonumber\\
&
\{2,2,2,2,2,1,0,1\}
,
\{0,0,1,1,1,1,0,0\}
,
\{1,0,1,2,2,2,2,2\}
,
\nonumber\\
&
\{1,1,0,0,1,1,0,0\}
,
\{2,1,1,2,2,1,0,1\}
,
\{1,1,2,2,1,0,1,2\}
,
\nonumber\\
&
\{1,0,1,2,2,2,1,1\}
,
\{2,1,0,1,2,2,1,1\}
,
\{2,1,0,1,2,2,2,2\}
,
\nonumber\\
&
\{2,2,1,1,2,1,0,1\}
,
\{1,1,1,1,1,0,0,1\}
,
\{1,1,0,0,1,1,1,1\}
,
\nonumber\\
&
\{2,2,1,1,1,2,1,0\}
,
\{1,0,0,1,1,0,0,1\}
,
\{2,1,0,1,1,2,2,1\}
,
\nonumber\\
&
\{1,1,0,0,0,1,1,0\}
,
\{1,0,0,1,1,1,0,0\}
,
\{0,0,1,1,1,1,1,1\}
,
\nonumber\\
&
\{1,0,0,1,1,1,1,1\}
,
\{0,0,1,1,0,0,1,1\}
\}
\, ,
\end{align}
associated with them.

Some of them can be calculated in a closed form. They are
\begin{alignat}{2}
&
{\rm Reg}_1 
&&
= \e^{2 \gamma_{\rm E}  \ep} (s_{23} s_{45})^{\ep} \Gamma^4 (\ep) \Gamma^2 (1 - \ep)
\, , \\
&
{\rm Reg}_5 
&&
= \e^{2 \gamma_{\rm E}  \ep} (s_{23} s_{34} )^{\ep} \frac{\Gamma^3 (\ep) \Gamma^2 (1 - \ep) \Gamma (2\ep)}{\Gamma (1 + \ep)}
\, , \\
&
{\rm Reg}_{12}
&&
= \e^{2 \gamma_{\rm E}  \ep} s_{23}^{\ep} \frac{\Gamma^3 (\ep) \Gamma^2 (- \ep) \Gamma (1 - \ep)}{\Gamma (- 2 \ep)}
\, ,
\end{alignat}
with others obtained from these via leg permutations. We will accommodate the latter with overall multiplicative prefactors such that
\begin{alignat}{2}
&
{\rm Reg}_2 
&&
= \left(\frac{s_{12}}{s_{23}} \right)^{\ep} {\rm Reg}_1
\, , \\
&
{\rm Reg}_6
&&
= \left(\frac{s_{12} s_{51}}{s_{23} s_{34}} \right)^{\ep} {\rm Reg}_5
\, , \\
&
{\rm Reg}_{17}
&&
= \left(\frac{s_{45}}{s_{23}} \right)^{\ep} {\rm Reg}_{12}
\, , \\
&
{\rm Reg}_{20}
&&
= \left(\frac{s_{12}}{s_{23}} \right)^{\ep} {\rm Reg}_{12}
\, .
\end{alignat}

All other regions are cast into the form
\begin{align}
{\rm Reg}_r
= 
C_i
\int \frac{d^{E_r} \bit{x}}{\mbox{GL(1)}} I_i (\bit{x})
\, , \\
\end{align}
where the dimension $E$ of the Feynman/Schwinger integrals can be ascertained by counting the number of Schwinger  parameters in each integrand.
We list below the overall coefficients 
\begin{alignat}{2}
\label{OverallFactorsApp}
&
C_3 
&&
= \e^{2 \gamma_{\rm E}  \ep} \Gamma (3 + 2 \ep)
\, , \\ 
&
C_4 
&&
= \frac{\e^{2 \gamma_{\rm E}  \ep}}{s_{45}} \Gamma (2 + 2\ep)
\, , \\ 
&
C_7 
&&
= \frac{\e^{2 \gamma_{\rm E}  \ep}}{s_{23} s_{45}}  \Gamma (2 \ep)
\, , \\
&
C_8 
&&
= \frac{\e^{2 \gamma_{\rm E}  \ep} s_{12}^{\ep} s_{51}}{s_{12} s_{23} s_{45}} \Gamma^3 (\ep) \Gamma (1 - \ep)
\, , \\
&
C_9
&&
= \frac{\e^{2 \gamma_{\rm E}  \ep}}{s_{12}} \Gamma (2\ep)
\, , \\
&
C_{10}
&&
= \frac{\e^{2 \gamma_{\rm E}  \ep} s_{51}}{s_{12} s_{45}} \Gamma (2 \ep)
\, , \\
&
C_{13}
&&
= \e^{2 \gamma_{\rm E}  \ep} \Gamma (2 + 2 \ep)
\, , \\
&
C_{15} 
&&
= \frac{\e^{2 \gamma_{\rm E}  \ep} s_{23}^{\ep} s_{34}}{s_{12} s_{23}} \Gamma^2 (\ep) \Gamma (1 - \ep) \Gamma (2 + \ep)
\, , \\
&
C_{18} 
&&
= \frac{\e^{2 \gamma_{\rm E}  \ep} s_{34}^{\ep}}{s_{23} s_{45}} \Gamma (\ep) \Gamma (1 - \ep) \Gamma (2 \ep)
\, , \\
&
C_{24} 
&&
= \frac{\e^{2 \gamma_{\rm E}  \ep} s_{51}}{s_{12}} \Gamma (2 + 2 \ep)
\, , \\
&
C_{26} 
&&
= \frac{\e^{2 \gamma_{\rm E}  \ep}}{s_{12} s_{23} s_{45}} \Gamma (2 \ep)
\, , \\
&
C_{28} 
&&
= \frac{\e^{2 \gamma_{\rm E}  \ep}}{s_{12}} \Gamma (2 \ep)
\, , \\
&
C_{29} 
&&
= \frac{\e^{2 \gamma_{\rm E}  \ep}}{s_{12} s_{23} s_{45}} \Gamma (2 \ep)
\, , \\
&
C_{31} 
&&
= \frac{\e^{2 \gamma_{\rm E}  \ep}}{s_{12} s_{23}} \Gamma (2 + 2 \ep)
\, .
\end{alignat}
The (yet to be gauge-fixed) integrands are\footnote{The reader should not be surprised that some of these integrands involve polynomials
in Schwinger parameters that are of mixed nature, i.e., with coefficients proportional to either $s_{jj+1}$ or unity. This is the artefact of
us setting the scale parameter of dimensional regularization to $\mu = 1$ such that $s_{jj+1}$ are effectively dimensionless. The monomials 
with unit coefficients in the MofR integrands stem from the ones with the off-shellness $m$ as a mass parameter in the original Symanzik 
polynomial $F$ \re{SymanzikPolys}. Upon the rescaling \re{MofRSchwingerRescaling}, the $m$-dependence gets converted to the 
$\mu$-dependence. Since we set the latter to one, it is no longer visible.}
\footnotesize
\begin{alignat}{2}
\label{Reg3Init}
&
I_3
&&
=
\partial_{x_9} [ U^{1 + 3 \ep} F^{-3 - 2 \ep} ]|_{x_9 = 0, m = 0}
\, , \\[2mm]
\label{Reg4Init}
&
I_4
&&
=
\frac{(x_7 + x_8)^{1 + 3 \ep}}{x_8}
(x_1 + x_2 + x_3 + x_4 + x_5)^{-1 + 3 \ep}
\big[
- 3 \ep
{\rm Denom}_4^{-2 - 2 \ep}
\\
&
&&
+
2 (1 + \ep)(x_1 + x_2 + x_3 + x_4 + x_5) [ (s_{34} x_2 + s_{51} x_3)(x_7 + x_8) + x_7 x_8]
{\rm Denom}_4^{-3 - 2 \ep}
\big]
\, , \nonumber\\[2mm]
\label{Reg7Init}
&
I_7
&&
=
\frac{(x_1+x_2)^{3 \ep} (x_6+x_7)^{3 \ep} (x_2 x_6 x_7 + x_1 x_6 x_7 + x_1 x_2 x_6 + x_1 x_2 x_7)^{-2 \ep}}{x_1 x_6 x_7 (s_{12} x_2+s_{45} x_1)}
\, , \\[2mm]
&
\label{Reg8Init}
I_8
&&
=
\frac{(x_6 x_7)^{- \ep} (x_6+x_7)^{2 \ep}}{x_6 (s_{23} x_6+s_{51} x_7)}
\, , \\[2mm]
\label{Reg9Init}
&
I_9
&&
=
\frac{x_2^{3 \ep} (x_5+x_6)^{3 \ep} (s_{23} x_3 x_5 x_6 + x_2 x_3 x_5 + x_2 x_3 x_6 + x_2 x_5 x_6)^{-2 \ep}}{x_2 x_5 (x_2+s_{23} x_3) (s_{12} x_5+s_{45} x_6)}
\, \\[2mm]
&
\label{Reg10Init}
I_{10}
&&
=
\frac{x_3 (x_3 + x_4)^{3 \ep} (x_6+x_7)^{3 \ep}(x_3 x_4 x_6 + x_3 x_4 x_7 + x_3 x_6 x_7 + x_4 x_6 x_7)^{- 2\ep} 
}{
x_4 x_6 (s_{23} x_3 +s_{45} x_4)(s_{23} x_3 x_6 + s_{45} x_4 x_6 + s_{51} x_3 x_7)}
\, , \\[2mm]
\label{Reg13Init}
&
I_{13}
&&
=
\frac{x_7^{\ep}}{x_7 (x_7 + s_{45} x_8)}
(x_1 + x_2 + x_3 + x_4 + x_5)^{-1 + 3 \ep}
\big[
- 3 \ep
{\rm Denom}_{13}^{-2 - 2 \ep}
\\
&
&&
+
2 (1 + \ep)(x_1 + x_2 + x_3 + x_4 + x_5) [ s_{34} x_2 + s_{51} x_3 + x_8]
{\rm Denom}_{13}^{-3 - 2 \ep}
\big]
\, , \nonumber\\[2mm]
\label{Reg15Init}
&
I_{15}
&&
=
(x_5 + x_6 + x_7 + x_8)^{2 \ep} (s_{12} x_5 x_8 + s_{34} x_5 x_7 + s_{45} x_6 x_8)^{- 2 \ep}
\, , \\[2mm]
\label{Reg18Init}
&
I_{18}
&&
=
\frac{(x_2 x_5)^{- \ep} (x_1+x_5)^{-1 - 2 \ep} (x_1 + x_2 + x_5)^{2 \ep}}{x_5 [s_{12} x_2+s_{45} (x_1 + x_5)]^{- 3 - 2 \ep}}
\, , \\[2mm]
\label{Reg24Init}
&
I_{24}
&&
=
\frac{x_3 (x_3+x_4)^{1 + 3 \ep} (x_5 + x_6 + x_7 + x_8 )^{3 \ep}}{x_4 (s_{23} x_3 + s_{45} x_4)}
\\
&
&&
\times
\left[
s_{23} x_3 x_5 x_6
+
s_{45} x_6 (x_4 x_5+x_3 x_8+x_4 x_8) 
+
s_{51} x_3 x_5 x_7
+
x_3 x_4 (x_5+x_6+x_7+x_8)
\right]^{-2 - 2 \ep}
\, , \\[2mm]
\label{Reg26Init}
&
I_{26}
&&
=
\frac{(x_2+x_3)^{3 \ep} (x_6+x_7)^{3 \ep} ( s_{34} x_2 + s_{51} x_3 )
}{
x_2 x_3 x_6
s_{23} x_3 x_6 + s_{34} x_2 x_7 + s_{51} x_3 x_7
}
(x_2 x_3 x_6+x_2 x_7 x_6+x_3 x_7 x_6+x_2 x_3 x_7)^{-2 \ep}
\, , \\[2mm]
\label{Reg28Init}
&
I_{28}
&&
=
\frac{x_3^{-2 \ep} (x_5 + x_8)}{x_5}
\frac{(x_3 x_5+x_4 x_5+x_8 x_5+x_3 x_8+x_4 x_8)^{1 + 3 \ep}
}{
s_{23} x_3 x_5 + s_{45} (x_4 x_5+x_8 x_5+x_3 x_8+x_4 x_8)
}
\frac{
[ x_5 x_8 + x_4 (x_5+x_8) ]^{-1 - 2 \ep}
}{
s_{23} x_3 (x_5+x_8) + s_{45} (x_4 x_5+x_8 x_5+x_4 x_8)
}
\, , \\
\label{Reg29Init}
&
I_{29}
&&
=
\frac{(x_2+x_3)^{3 \ep} (x_7+x_8)^{3 \ep} ( s_{34} x_2 + s_{51} x_3)}{x_2 x_3 x_8}
\frac{
\left(x_2 x_3 x_7+x_2 x_8 x_7+x_3 x_8 x_7+x_2 x_3 x_8\right)^{-2 \ep}
}{s_{12} x_2 x_8 + s_{34} x_2 x_7 + s_{51} x_3 x_7
}
\, , \\
\label{Reg31Init}
&
I_{31}
&&
=
\frac{
(x_2+x_3)^{1 + 3 \ep} (x_5+x_6+x_7+x_8)^{3\ep} (s_{34} x_2 + s_{51} x_3)
}{
x_2 x_3}
\\
&
&&
\times\!
[
s_{12} x_2 x_5 x_8
+
s_{23} x_3 x_5 x_6
+
s_{34} x_2 x_5 x_7
+
s_{45} x_6 x_8 (x_2+x_3)
+
s_{51} x_3 x_5 x_7
+
x_2 x_3 (x_5+x_6+x_7+x_8)
]^{-2-2 \ep} 
\!\! . \nonumber\\[2mm]
\end{alignat}
\normalsize
Here, the hard region 3, corresponding to the region vector $\bit{r}_3 = \{ 0,0,0,0,0,0,0,0\}$, is defined by the original integrand
\re{FeynmanIntegralInput} with $m$ identically set to zero. To make these formulas more readable, we introduced shorthand 
notations for the lengthy denominators involved
\begin{align}
&
{\rm Denom}_4 
= 
s_{12} x_2 (x_4 x_7+x_4 x_8+x_5 x_8 ) 
+ 
s_{23} x_1 x_3 (x_7+x_8)
\\
&\quad
+ 
s_{34} x_2 x_5 x_7
+
s_{45} x_1 (x_4 x_7+x_4 x_8+x_5 x_8)
+
s_{51} x_3 x_5 x_7
+
x_7 x_8
(x_1+x_2+x_3+x_4+x_5)
\, , \nonumber\\
&
{\rm Denom}_{13} 
= 
s_{12} x_2 x_4 + s_{23} x_1 x_3 + s_{34} x_2 x_5 + s_{45} x_1 x_4 + s_{51} x_5 x_3
+ 
(x_1+x_2+x_3+x_4+x_5) x_8
\, .
\end{align}

We displayed above only expressions for functionally different regions. Those not shown are again obtained from these by exchanges
of the Mandelstam-like variables and factor multiplications. Namely,
\begin{alignat}{2}
&
{\rm Reg}_{11} 
&&
= \frac{s_{51}}{s_{12}} \left(\frac{s_{45}}{s_{51}} \right)^{\ep}  {\rm Reg}_8|_{s_{23} \leftrightarrow s_{45}, s_{51} \leftrightarrow s_{12}}
\, , \\
&
{\rm Reg}_{19} 
&&
= {\rm Reg}_8|_{s_{12} \leftrightarrow s_{23}, s_{34} \leftrightarrow s_{51}}
\, , \\
&
{\rm Reg}_{22} 
&&
= 
\frac{s_{34}}{s_{23}} \left(\frac{s_{45}}{s_{34}} \right)^{\ep}
{\rm Reg}_8|_{s_{12} \to s_{34}, s_{23} \to s_{45}, s_{45} \to s_{12}, s_{51} \to s_{23}}
\, , \\
&
{\rm Reg}_{27} 
&&
= 
\frac{s_{51}}{s_{45}} \frac{\Gamma (2 \ep)}{\Gamma (\ep) \Gamma (1+\ep)}
{\rm Reg}_8|_{s_{45} \leftrightarrow s_{51}}
\, , \\
&
{\rm Reg}_{25} 
&&
= 
{\rm Reg}_{18}|_{s_{12} \leftrightarrow s_{23}, s_{34} \leftrightarrow s_{51}}
\, , \\
&
{\rm Reg}_{16} 
&&
= 
{\rm Reg}_{7}|_{s_{12} \leftrightarrow s_{23}}
\, , \\
&
{\rm Reg}_{21} 
&&
= 
{\rm Reg}_{15}|_{s_{12} \leftrightarrow s_{23}, s_{34} \leftrightarrow s_{51}}
\, , \\
&
{\rm Reg}_{14} 
&&
= 
{\rm Reg}_{10}|_{s_{12} \leftrightarrow s_{23}, s_{34} \leftrightarrow s_{51}}
\, , \\
&
{\rm Reg}_{23} 
&&
= 
{\rm Reg}_{4}|_{s_{12} \leftrightarrow s_{23}, s_{34} \leftrightarrow s_{51}}
\, , \\
&
{\rm Reg}_{30} 
&&
= 
{\rm Reg}_{24}|_{s_{12} \leftrightarrow s_{23}, s_{34} \leftrightarrow s_{51}}
\, , \\
&
{\rm Reg}_{32} 
&&
= 
{\rm Reg}_{28}|_{s_{12} \leftrightarrow s_{23}, s_{34} \leftrightarrow s_{51}}
\, .
\end{alignat}



\begin{thebibliography}{100}

\bibitem{Selivanov:1999ie}
K.~G. Selivanov, \emph{{An Infinite set of tree amplitudes in
  Higgs-Yang-Mills}},
  \href{http://dx.doi.org/10.1016/S0370-2693(99)00760-1}{\emph{Phys. Lett. B}
  {\bf 460} (1999) 116--118}, [\href{https://arxiv.org/abs/hep-th/9906001}{{\tt
  hep-th/9906001}}].

\bibitem{Alday:2009zm}
L.~F. Alday, J.~M. Henn, J.~Plefka and T.~Schuster, \emph{{Scattering into the
  fifth dimension of N=4 super Yang-Mills}},
  \href{http://dx.doi.org/10.1007/JHEP01(2010)077}{\emph{JHEP} {\bf 01} (2010)
  077}, [\href{https://arxiv.org/abs/0908.0684}{{\tt 0908.0684}}].

\bibitem{Boels:2010mj}
R.~H. Boels, \emph{{No triangles on the moduli space of maximally
  supersymmetric gauge theory}},
  \href{http://dx.doi.org/10.1007/JHEP05(2010)046}{\emph{JHEP} {\bf 05} (2010)
  046}, [\href{https://arxiv.org/abs/1003.2989}{{\tt 1003.2989}}].

\bibitem{Craig:2011ws}
N.~Craig, H.~Elvang, M.~Kiermaier and T.~Slatyer, \emph{{Massive amplitudes on
  the Coulomb branch of N=4 SYM}},
  \href{http://dx.doi.org/10.1007/JHEP12(2011)097}{\emph{JHEP} {\bf 12} (2011)
  097}, [\href{https://arxiv.org/abs/1104.2050}{{\tt 1104.2050}}].

\bibitem{Caron-Huot:2021usw}
S.~Caron-Huot and F.~Coronado, \emph{{Ten dimensional symmetry of $ \mathcal{N}
  $ = 4 SYM correlators}},
  \href{http://dx.doi.org/10.1007/JHEP03(2022)151}{\emph{JHEP} {\bf 03} (2022)
  151}, [\href{https://arxiv.org/abs/2106.03892}{{\tt 2106.03892}}].

\bibitem{Bork:2022vat}
L.~V. Bork, N.~B. Muzhichkov and E.~S. Sozinov, \emph{{Infrared properties of
  five-point massive amplitudes in $ \mathcal{N} $ = 4 SYM on the Coulomb
  branch}}, \href{http://dx.doi.org/10.1007/JHEP08(2022)173}{\emph{JHEP} {\bf
  08} (2022) 173}, [\href{https://arxiv.org/abs/2201.08762}{{\tt 2201.08762}}].

\bibitem{Belitsky:2022itf}
A.~V. Belitsky, L.~V. Bork, A.~F. Pikelner and V.~A. Smirnov, \emph{{Exact Off
  Shell Sudakov Form Factor in N=4 Supersymmetric Yang-Mills Theory}},
  \href{http://dx.doi.org/10.1103/PhysRevLett.130.091605}{\emph{Phys. Rev.
  Lett.} {\bf 130} (2023) 091605},
  [\href{https://arxiv.org/abs/2209.09263}{{\tt 2209.09263}}].

\bibitem{Belitsky:2023ssv}
A.~V. Belitsky, L.~V. Bork and V.~A. Smirnov, \emph{{Off-shell form factor in $
  \mathcal{N} $=4 sYM at three loops}},
  \href{http://dx.doi.org/10.1007/JHEP11(2023)111}{\emph{JHEP} {\bf 11} (2023)
  111}, [\href{https://arxiv.org/abs/2306.16859}{{\tt 2306.16859}}].

\bibitem{Belitsky:2024agy}
A.~V. Belitsky, L.~V. Bork, J.~M. Grumski-Flores and V.~A. Smirnov,
  \emph{{Three-leg form factor on Coulomb branch}},
  \href{http://dx.doi.org/10.1007/JHEP11(2024)169}{\emph{JHEP} {\bf 11} (2024)
  169}, [\href{https://arxiv.org/abs/2402.18475}{{\tt 2402.18475}}].

\bibitem{Belitsky:2024dcf}
A.~V. Belitsky and L.~V. Bork, \emph{{Off-shell minimal form factors}},
  \href{http://dx.doi.org/10.1007/JHEP07(2025)231}{\emph{JHEP} {\bf 07} (2025)
  231}, [\href{https://arxiv.org/abs/2411.16941}{{\tt 2411.16941}}].

\bibitem{Belitsky:2025bez}
A.~V. Belitsky and V.~A. Smirnov, \emph{{Off-shell form factor: factorization
  is violated}},  \href{https://arxiv.org/abs/2505.22595}{{\tt 2505.22595}}.

\bibitem{Polyakov:1980ca}
A.~M. Polyakov, \emph{{Gauge Fields as Rings of Glue}},
  \href{http://dx.doi.org/10.1016/0550-3213(80)90507-6}{\emph{Nucl. Phys. B}
  {\bf 164} (1980) 171--188}.

\bibitem{Korchemsky:1987wg}
G.~P. Korchemsky and A.~V. Radyushkin, \emph{{Renormalization of the Wilson
  Loops Beyond the Leading Order}},
  \href{http://dx.doi.org/10.1016/0550-3213(87)90277-X}{\emph{Nucl. Phys. B}
  {\bf 283} (1987) 342--364}.

\bibitem{Coronado:2018cxj}
F.~Coronado, \emph{{Bootstrapping the Simplest Correlator in Planar $\mathcal N
  = 4$ Supersymmetric Yang-Mills Theory to All Loops}},
  \href{http://dx.doi.org/10.1103/PhysRevLett.124.171601}{\emph{Phys. Rev.
  Lett.} {\bf 124} (2020) 171601},
  [\href{https://arxiv.org/abs/1811.03282}{{\tt 1811.03282}}].

\bibitem{Belitsky:2019fan}
A.~V. Belitsky and G.~P. Korchemsky, \emph{{Exact null octagon}},
  \href{http://dx.doi.org/10.1007/JHEP05(2020)070}{\emph{JHEP} {\bf 05} (2020)
  070}, [\href{https://arxiv.org/abs/1907.13131}{{\tt 1907.13131}}].

\bibitem{Belitsky:2020qzm}
A.~V. Belitsky, \emph{{Null octagon from Deift-Zhou steepest descent}},
  \href{http://dx.doi.org/10.1016/j.nuclphysb.2022.115844}{\emph{Nucl. Phys. B}
  {\bf 980} (2022) 115844}, [\href{https://arxiv.org/abs/2012.10446}{{\tt
  2012.10446}}].

\bibitem{Anastasiou:2003kj}
C.~Anastasiou, Z.~Bern, L.~J. Dixon and D.~A. Kosower, \emph{{Planar amplitudes
  in maximally supersymmetric Yang-Mills theory}},
  \href{http://dx.doi.org/10.1103/PhysRevLett.91.251602}{\emph{Phys. Rev.
  Lett.} {\bf 91} (2003) 251602},
  [\href{https://arxiv.org/abs/hep-th/0309040}{{\tt hep-th/0309040}}].

\bibitem{Bern:2005iz}
Z.~Bern, L.~J. Dixon and V.~A. Smirnov, \emph{{Iteration of planar amplitudes
  in maximally supersymmetric Yang-Mills theory at three loops and beyond}},
  \href{http://dx.doi.org/10.1103/PhysRevD.72.085001}{\emph{Phys. Rev. D} {\bf
  72} (2005) 085001}, [\href{https://arxiv.org/abs/hep-th/0505205}{{\tt
  hep-th/0505205}}].

\bibitem{He:2025vqt}
S.~He, X.~Jiang, J.~Liu and Y.-Q. Zhang, \emph{{Notes on conformal integrals:
  Coulomb branch amplitudes, magic identities and bootstrap}},
  \href{https://arxiv.org/abs/2502.08871}{{\tt 2502.08871}}.

\bibitem{He:2022ctv}
S.~He, Z.~Li, R.~Ma, Z.~Wu, Q.~Yang and Y.~Zhang, \emph{{A study of Feynman
  integrals with uniform transcendental weights and their symbology}},
  \href{http://dx.doi.org/10.1007/JHEP10(2022)165}{\emph{JHEP} {\bf 10} (2022)
  165}, [\href{https://arxiv.org/abs/2206.04609}{{\tt 2206.04609}}].

\bibitem{Belitsky:2023gba}
A.~V. Belitsky and V.~A. Smirnov, \emph{{Near mass-shell double boxes}},
  \href{http://dx.doi.org/10.1007/JHEP05(2024)155}{\emph{JHEP} {\bf 05} (2024)
  155}, [\href{https://arxiv.org/abs/2312.00641}{{\tt 2312.00641}}].

\bibitem{Belitsky:2024rwv}
A.~V. Belitsky, \emph{{Collinear anatomy}},
  \href{http://dx.doi.org/10.1007/JHEP05(2025)117}{\emph{JHEP} {\bf 05} (2025)
  117}, [\href{https://arxiv.org/abs/2412.11886}{{\tt 2412.11886}}].

\bibitem{Kotikov:1990kg}
A.~V. Kotikov, \emph{{Differential equations method: New technique for massive
  Feynman diagrams calculation}},
  \href{http://dx.doi.org/10.1016/0370-2693(91)90413-K}{\emph{Phys. Lett. B}
  {\bf 254} (1991) 158--164}.

\bibitem{Gehrmann:1999as}
T.~Gehrmann and E.~Remiddi, \emph{{Differential equations for two loop four
  point functions}},
  \href{http://dx.doi.org/10.1016/S0550-3213(00)00223-6}{\emph{Nucl. Phys. B}
  {\bf 580} (2000) 485--518}, [\href{https://arxiv.org/abs/hep-ph/9912329}{{\tt
  hep-ph/9912329}}].

\bibitem{Henn:2013pwa}
J.~M. Henn, \emph{{Multiloop integrals in dimensional regularization made
  simple}}, \href{http://dx.doi.org/10.1103/PhysRevLett.110.251601}{\emph{Phys.
  Rev. Lett.} {\bf 110} (2013) 251601},
  [\href{https://arxiv.org/abs/1304.1806}{{\tt 1304.1806}}].

\bibitem{Belitsky:2024jhe}
A.~V. Belitsky, A.~A. Kokosinskaya, A.~V. Smirnov, V.~V. Voevodin and M.~Zeng,
  \emph{{Efficient Reduction of Feynman Integrals on Supercomputers}},
  \href{http://dx.doi.org/10.1134/S1995080224603709}{\emph{Lobachevskii J.
  Math.} {\bf 45} (2024) 2984--2994},
  [\href{https://arxiv.org/abs/2402.07499}{{\tt 2402.07499}}].

\bibitem{Panzer:2014gra}
E.~Panzer, \emph{{On hyperlogarithms and Feynman integrals with divergences and
  many scales}}, \href{http://dx.doi.org/10.1007/JHEP03(2014)071}{\emph{JHEP}
  {\bf 03} (2014) 071}, [\href{https://arxiv.org/abs/1401.4361}{{\tt
  1401.4361}}].

\bibitem{Beneke:1997zp}
M.~Beneke and V.~A. Smirnov, \emph{{Asymptotic expansion of Feynman integrals
  near threshold}},
  \href{http://dx.doi.org/10.1016/S0550-3213(98)00138-2}{\emph{Nucl. Phys. B}
  {\bf 522} (1998) 321--344}, [\href{https://arxiv.org/abs/hep-ph/9711391}{{\tt
  hep-ph/9711391}}].

\bibitem{Smirnov:1999gc}
V.~A. Smirnov, \emph{{Analytical result for dimensionally regularized massless
  on shell double box}},
  \href{http://dx.doi.org/10.1016/S0370-2693(99)00777-7}{\emph{Phys. Lett. B}
  {\bf 460} (1999) 397--404}, [\href{https://arxiv.org/abs/hep-ph/9905323}{{\tt
  hep-ph/9905323}}].

\bibitem{Tausk:1999vh}
J.~B. Tausk, \emph{{Nonplanar massless two loop Feynman diagrams with four
  on-shell legs}},
  \href{http://dx.doi.org/10.1016/S0370-2693(99)01277-0}{\emph{Phys. Lett. B}
  {\bf 469} (1999) 225--234}, [\href{https://arxiv.org/abs/hep-ph/9909506}{{\tt
  hep-ph/9909506}}].

\bibitem{Czakon:2005rk}
M.~Czakon, \emph{{Automatized analytic continuation of Mellin-Barnes
  integrals}}, \href{http://dx.doi.org/10.1016/j.cpc.2006.07.002}{\emph{Comput.
  Phys. Commun.} {\bf 175} (2006) 559--571},
  [\href{https://arxiv.org/abs/hep-ph/0511200}{{\tt hep-ph/0511200}}].

\bibitem{Smirnov:2009up}
A.~V. Smirnov and V.~A. Smirnov, \emph{{On the Resolution of Singularities of
  Multiple Mellin-Barnes Integrals}},
  \href{http://dx.doi.org/10.1140/epjc/s10052-009-1039-6}{\emph{Eur. Phys. J.
  C} {\bf 62} (2009) 445--449}, [\href{https://arxiv.org/abs/0901.0386}{{\tt
  0901.0386}}].

\bibitem{Binoth:2000ps}
T.~Binoth and G.~Heinrich, \emph{{An automatized algorithm to compute infrared
  divergent multi-loop integrals}},
  \href{http://dx.doi.org/10.1016/S0550-3213(00)00429-6}{\emph{Nucl. Phys. B}
  {\bf 585} (2000) 741--759}, [\href{https://arxiv.org/abs/hep-ph/0004013}{{\tt
  hep-ph/0004013}}].

\bibitem{Bogner:2007cr}
C.~Bogner and S.~Weinzierl, \emph{{Resolution of singularities for multi-loop
  integrals}}, \href{http://dx.doi.org/10.1016/j.cpc.2007.11.012}{\emph{Comput.
  Phys. Commun.} {\bf 178} (2008) 596--610},
  [\href{https://arxiv.org/abs/0709.4092}{{\tt 0709.4092}}].

\bibitem{Kaneko:2009qx}
T.~Kaneko and T.~Ueda, \emph{{A Geometric method of sector decomposition}},
  \href{http://dx.doi.org/10.1016/j.cpc.2010.04.001}{\emph{Comput. Phys.
  Commun.} {\bf 181} (2010) 1352--1361},
  [\href{https://arxiv.org/abs/0908.2897}{{\tt 0908.2897}}].

\bibitem{Smirnov:2021rhf}
A.~V. Smirnov, N.~D. Shapurov and L.~I. Vysotsky, \emph{{FIESTA5: Numerical
  high-performance Feynman integral evaluation}},
  \href{http://dx.doi.org/10.1016/j.cpc.2022.108386}{\emph{Comput. Phys.
  Commun.} {\bf 277} (2022) 108386},
  [\href{https://arxiv.org/abs/2110.11660}{{\tt 2110.11660}}].

\bibitem{Nilsson}
L.~Nilsson and M.~Passare, \emph{{Mellin transforms of multivariate rational
  functions}}, \href{http://dx.doi.org/10.48550/arXiv.1010.5060}{\emph{arXiv
  e-prints} (Oct., 2010) arXiv:1010.5060},
  [\href{https://arxiv.org/abs/1010.5060}{{\tt 1010.5060}}].

\bibitem{Berkesch}
C.~Berkesch, J.~Forsg{\aa}rd and M.~Passare, \emph{Euler-mellin integrals and
  a-hypergeometric functions},
  \href{http://dx.doi.org/10.1307/mmj/1395234361}{\emph{Michigan Mathematical
  Journal} {\bf 63} (Mar., 2014) 101--123}.

\bibitem{Speer:1968qxh}
E.~R. Speer, \emph{{Analytic Renormalization}},
  \href{http://dx.doi.org/10.1063/1.1664729}{\emph{J. Math. Phys.} {\bf 9}
  (1968) 1404}.

\bibitem{Speer:1971fub}
E.~R. Speer, \emph{{On the structure of analytic renormalization}},
  \href{http://dx.doi.org/10.1007/BF01877594}{\emph{Commun. Math. Phys.} {\bf
  23} (1971) 23--36}.

\bibitem{Breitenlohner:1975hg}
P.~Breitenlohner and D.~Maison, \emph{{Dimensionally Renormalized Green's
  Functions for Theories with Massless Particles. 1.}},
  \href{http://dx.doi.org/10.1007/BF01609070}{\emph{Commun. Math. Phys.} {\bf
  52} (1977) 39}.

\bibitem{Breitenlohner:1976te}
P.~Breitenlohner and D.~Maison, \emph{{Dimensionally Renormalized Green's
  Functions for Theories with Massless Particles. 2.}},
  \href{http://dx.doi.org/10.1007/BF01609071}{\emph{Commun. Math. Phys.} {\bf
  52} (1977) 55}.

\bibitem{tHooft:1972tcz}
G.~'t~Hooft and M.~J.~G. Veltman, \emph{{Regularization and Renormalization of
  Gauge Fields}},
  \href{http://dx.doi.org/10.1016/0550-3213(72)90279-9}{\emph{Nucl. Phys. B}
  {\bf 44} (1972) 189--213}.

\bibitem{Salvatori:2024nva}
G.~Salvatori, \emph{{The Tropical Geometry of Subtraction Schemes}},
  \href{https://arxiv.org/abs/2406.14606}{{\tt 2406.14606}}.

\bibitem{Cheng:1987ga}
H.~Cheng and T.~T. Wu, \emph{{Expanding protons: scatterinf at high energies}}.
\newblock MIT Press, 1987.

\bibitem{Smirnov:1999bza}
V.~A. Smirnov, \emph{{Problems of the strategy of regions}},
  \href{http://dx.doi.org/10.1016/S0370-2693(99)01061-8}{\emph{Phys. Lett. B}
  {\bf 465} (1999) 226--234}, [\href{https://arxiv.org/abs/hep-ph/9907471}{{\tt
  hep-ph/9907471}}].

\bibitem{Pak:2010pt}
A.~Pak and A.~Smirnov, \emph{{Geometric approach to asymptotic expansion of
  Feynman integrals}},
  \href{http://dx.doi.org/10.1140/epjc/s10052-011-1626-1}{\emph{Eur. Phys. J.
  C} {\bf 71} (2011) 1626}, [\href{https://arxiv.org/abs/1011.4863}{{\tt
  1011.4863}}].

\bibitem{Belitsky:2024yag}
A.~V. Belitsky, L.~V. Bork and V.~A. Smirnov, \emph{{Pinching Sudakov}},
  \href{http://dx.doi.org/10.1007/JHEP05(2025)237}{\emph{JHEP} {\bf 05} (2025)
  237}, [\href{https://arxiv.org/abs/2409.05945}{{\tt 2409.05945}}].

\bibitem{Barber:1996lmi}
C.~B. Barber, D.~P. Dobkin and H.~Huhdanpaa, \emph{{The quickhull algorithm for
  convex hulls}}, \href{http://dx.doi.org/10.1145/235815.235821}{\emph{ACM
  Trans. Math. Software} {\bf 22} (1996) 469--483}.

\bibitem{Arkani-Hamed:2022cqe}
N.~Arkani-Hamed, A.~Hillman and S.~Mizera, \emph{{Feynman polytopes and the
  tropical geometry of UV and IR divergences}},
  \href{http://dx.doi.org/10.1103/PhysRevD.105.125013}{\emph{Phys. Rev. D} {\bf
  105} (2022) 125013}, [\href{https://arxiv.org/abs/2202.12296}{{\tt
  2202.12296}}].

\bibitem{Rau2017}
J.~Rau, ``{\it A First Expedition to Tropical Geometry}.''
  \url{https://math.uniandes.edu.co/~j.rau/downloads/FirstExpedition.pdf},
  2017.

\bibitem{2015arXiv150205950B}
E.~{Brugall{\'e}}, I.~{Itenberg}, G.~{Mikhalkin} and K.~{Shaw}, \emph{{Brief
  introduction to tropical geometry}},
  \href{http://dx.doi.org/10.48550/arXiv.1502.05950}{\emph{arXiv} (Feb., 2015)
  arXiv:1502.05950}, [\href{https://arxiv.org/abs/1502.05950}{{\tt
  1502.05950}}].

\bibitem{MikhalkinRau2019}
G.~Mikhalkin and J.~Rau, ``{\it Tropical Geometry}.''
  \url{https://math.uniandes.edu.co/~j.rau/downloads/main.pdf}, 2019.

\bibitem{z-lop-95}
G.~M. Ziegler, \emph{Lectures on polytopes}.
\newblock Springer-Verlag, New York, 1995.

\bibitem{MartinEtAl1999}
M.~Henk, R.-G. Jürgen and G.~Ziegler, ``Basic properties of convex polytopes,
  in handbook of discrete and computational geometry.''
  \url{https://page.math.tu-berlin.de/~henk/preprints/henk richter-gebert
  ziegler&basic properties of convex polytopes.pdf}, 1999.

\bibitem{Oda1988}
T.~Oda, \emph{Convex bodies and algebraic geometry}.
\newblock Springer, 1988.

\bibitem{Borowka:2017idc}
S.~Borowka, G.~Heinrich, S.~Jahn, S.~P. Jones, M.~Kerner, J.~Schlenk et~al.,
  \emph{{pySecDec: A toolbox for the numerical evaluation of multi-scale
  integrals}}, \href{http://dx.doi.org/10.1016/j.cpc.2017.09.015}{\emph{Comput.
  Phys. Commun.} {\bf 222} (2018) 313--326},
  [\href{https://arxiv.org/abs/1703.09692}{{\tt 1703.09692}}].

\bibitem{Bogoliubov:1957gp}
N.~N. Bogoliubov and O.~S. Parasiuk, \emph{{On the Multiplication of the causal
  function in the quantum theory of fields}},
  \href{http://dx.doi.org/10.1007/BF02392399}{\emph{Acta Math.} {\bf 97} (1957)
  227--266}.

\bibitem{Hepp:1966eg}
K.~Hepp, \emph{{Proof of the Bogolyubov-Parasiuk theorem on renormalization}},
  \href{http://dx.doi.org/10.1007/BF01773358}{\emph{Commun. Math. Phys.} {\bf
  2} (1966) 301--326}.

\bibitem{Zimmermann:1969jj}
W.~Zimmermann, \emph{{Convergence of Bogolyubov's method of renormalization in
  momentum space}}, \href{http://dx.doi.org/10.1007/BF01645676}{\emph{Commun.
  Math. Phys.} {\bf 15} (1969) 208--234}.

\bibitem{Lowenstein:1975pd}
J.~H. Lowenstein, \emph{{Auxiliary Mass Formulation of the Pure Yang-Mills
  Model}}, \href{http://dx.doi.org/10.1016/0550-3213(75)90578-7}{\emph{Nucl.
  Phys. B} {\bf 96} (1975) 189--208}.

\bibitem{Hillman:2023ezp}
A.~Hillman, \emph{{A Subtraction Scheme for Feynman Integrals}},
  \href{https://arxiv.org/abs/2311.03439}{{\tt 2311.03439}}.

\bibitem{Hillman:2023vas}
A.~Hillman, \emph{{On the Infrared and Ultraviolet Behavior of Scattering
  Amplitudes and Wavefunctions}}.
\newblock PhD thesis, Princeton U., 2023.

\bibitem{Brown:2019wna}
F.~Brown and C.~Dupont, \emph{{Single-valued integration and superstring
  amplitudes in genus zero}},
  \href{http://dx.doi.org/10.1007/s00220-021-03969-4}{\emph{Commun. Math.
  Phys.} {\bf 382} (2021) 815--874},
  [\href{https://arxiv.org/abs/1910.01107}{{\tt 1910.01107}}].

\bibitem{Arkani-Hamed:2019mrd}
N.~Arkani-Hamed, S.~He and T.~Lam, \emph{{Stringy canonical forms}},
  \href{http://dx.doi.org/10.1007/JHEP02(2021)069}{\emph{JHEP} {\bf 02} (2021)
  069}, [\href{https://arxiv.org/abs/1912.08707}{{\tt 1912.08707}}].

\bibitem{Frellesvig:2016ske}
H.~Frellesvig, D.~Tommasini and C.~Wever, \emph{{On the reduction of
  generalized polylogarithms to $\text{Li}_n$ and $\text{Li}_{2,2}$ and on the
  evaluation thereof}},
  \href{http://dx.doi.org/10.1007/JHEP03(2016)189}{\emph{JHEP} {\bf 03} (2016)
  189}, [\href{https://arxiv.org/abs/1601.02649}{{\tt 1601.02649}}].

\bibitem{Schultka:2018nrs}
K.~Schultka, \emph{{Toric geometry and regularization of Feynman integrals}},
  \href{https://arxiv.org/abs/1806.01086}{{\tt 1806.01086}}.

\bibitem{Mathematica}
Wolfram, ``Mathematica, {V}ersion 14.3.''

\bibitem{polymake}
E.~Gawrilow and M.~Joswig, \emph{{Polymake: a framework for analyzing convex
  polytopes}}, {\emph{Polytopes --- Combinatorics and Computation} {\bf 29}
  (2000) 43--73}.

\bibitem{irobotquote}
A.~Proyas, \emph{I, Robot}.
\newblock Twentieth Century Fox, 2004.

\bibitem{Gehrmann:2018yef}
T.~Gehrmann, J.~M. Henn and N.~A. Lo~Presti, \emph{{Pentagon functions for
  massless planar scattering amplitudes}},
  \href{http://dx.doi.org/10.1007/JHEP10(2018)103}{\emph{JHEP} {\bf 10} (2018)
  103}, [\href{https://arxiv.org/abs/1807.09812}{{\tt 1807.09812}}].

\bibitem{Papadopoulos:2015jft}
C.~G. Papadopoulos, D.~Tommasini and C.~Wever, \emph{{The Pentabox Master
  Integrals with the Simplified Differential Equations approach}},
  \href{http://dx.doi.org/10.1007/JHEP04(2016)078}{\emph{JHEP} {\bf 04} (2016)
  078}, [\href{https://arxiv.org/abs/1511.09404}{{\tt 1511.09404}}].

\bibitem{Chicherin:2020oor}
D.~Chicherin and V.~Sotnikov, \emph{{Pentagon Functions for Scattering of Five
  Massless Particles}},
  \href{http://dx.doi.org/10.1007/JHEP12(2020)167}{\emph{JHEP} {\bf 20} (2020)
  167}, [\href{https://arxiv.org/abs/2009.07803}{{\tt 2009.07803}}].

\bibitem{Duhr:2019tlz}
C.~Duhr and F.~Dulat, \emph{{PolyLogTools {\textemdash} polylogs for the
  masses}}, \href{http://dx.doi.org/10.1007/JHEP08(2019)135}{\emph{JHEP} {\bf
  08} (2019) 135}, [\href{https://arxiv.org/abs/1904.07279}{{\tt 1904.07279}}].

\bibitem{Bern:2008ap}
Z.~Bern, L.~J. Dixon, D.~A. Kosower, R.~Roiban, M.~Spradlin, C.~Vergu et~al.,
  \emph{{The Two-Loop Six-Gluon MHV Amplitude in Maximally Supersymmetric
  Yang-Mills Theory}},
  \href{http://dx.doi.org/10.1103/PhysRevD.78.045007}{\emph{Phys. Rev. D} {\bf
  78} (2008) 045007}, [\href{https://arxiv.org/abs/0803.1465}{{\tt
  0803.1465}}].

\bibitem{Drummond:2008aq}
J.~M. Drummond, J.~Henn, G.~P. Korchemsky and E.~Sokatchev, \emph{{Hexagon
  Wilson loop = six-gluon MHV amplitude}},
  \href{http://dx.doi.org/10.1016/j.nuclphysb.2009.02.015}{\emph{Nucl. Phys. B}
  {\bf 815} (2009) 142--173}, [\href{https://arxiv.org/abs/0803.1466}{{\tt
  0803.1466}}].

\bibitem{Dixon:2011nj}
L.~J. Dixon, J.~M. Drummond and J.~M. Henn, \emph{{Analytic result for the
  two-loop six-point NMHV amplitude in N=4 super Yang-Mills theory}},
  \href{http://dx.doi.org/10.1007/JHEP01(2012)024}{\emph{JHEP} {\bf 01} (2012)
  024}, [\href{https://arxiv.org/abs/1111.1704}{{\tt 1111.1704}}].

\bibitem{Brandhuber:2012vm}
A.~Brandhuber, G.~Travaglini and G.~Yang, \emph{{Analytic two-loop form factors
  in N=4 SYM}}, \href{http://dx.doi.org/10.1007/JHEP05(2012)082}{\emph{JHEP}
  {\bf 05} (2012) 082}, [\href{https://arxiv.org/abs/1201.4170}{{\tt
  1201.4170}}].

\bibitem{Brandhuber:2014ica}
A.~Brandhuber, B.~Penante, G.~Travaglini and C.~Wen, \emph{{The last of the
  simple remainders}},
  \href{http://dx.doi.org/10.1007/JHEP08(2014)100}{\emph{JHEP} {\bf 08} (2014)
  100}, [\href{https://arxiv.org/abs/1406.1443}{{\tt 1406.1443}}].

\bibitem{Belitsky:2025pnw}
A.~V. Belitsky and V.~A. Smirnov, \emph{{Collinear bootstrap for N=(1,1) sYM}},
   \href{https://arxiv.org/abs/2503.21915}{{\tt 2503.21915}}.

\end{thebibliography}
\end{document}